\documentclass[reprint,aip,pof,onecolumn]{revtex4-1}  

\usepackage{graphicx}
\usepackage{dcolumn}
\usepackage{bm}
 \usepackage{color}
\usepackage{float}

\begin{document}

\title{The role of the `monopole' instability in the evolution of 2D turbulent free shear layers}

\author{Saikishan Suryanarayanan}
 \email{saikishan.suryanarayanan@gmail.com}	
\affiliation{Jawaharlal Nehru Centre for Advanced Scientific Research, Bengaluru, India}%
\affiliation{Present address: The University of Texas at Austin, Austin, Texas 78751, United States of America}%
\author{Garry L. Brown}%
 \email{glb1873@msn.com }
\affiliation{Princeton University, Princeton, New Jersey 08544, USA, United States of America}%
\author{Roddam Narasimha}%
 \email{roddam@jncasr.ac.in}
\affiliation{Jawaharlal Nehru Centre for Advanced Scientific Research, Bengaluru, India}%

\date{\today}

\begin{abstract}
The role of instability in the growth of a 2D, temporally evolving, `turbulent' free shear layer is analyzed using vortex-gas simulations that condense all dynamics into the kinematics of the Biot-Savart relation.  The initial evolution of perturbations in a constant-vorticity layer is found to be in accurate agreement with the linear stability theory of Rayleigh. There is then a stage of non-universal evolution of coherent structures that is closely approximated not by Rayleigh stability theory, but by the Karman-Rubach-Lamb linear instability of monopoles, until the neighboring coherent structures merge. After several mergers, the layer evolves eventually to a self-preserving reverse cascade, characterized by a universal spread rate found by Suryanarayanan \textit{et al.} (\textit{Phys.Rev.E} 89, 013009, 2014) and a universal value of the ratio of dominant spacing of structures ($\Lambda_f$) to the layer thickness ($\delta_\omega$). In this universal, self-preserving state, the local amplification of perturbation amplitudes is accurately predicted by Rayleigh theory for the locally existing `base' flow. The model of Morris \textit{et al.} (\textit{Proc.Roy.Soc. A} 431, 219-243, 1990.), which computes the growth of the layer by balancing the energy lost by the mean flow with the energy gain of the perturbation modes (computed from an application of Rayleigh theory), is shown, however, to provide a non-universal asymptotic state with initial condition dependent spread-rate and spectra. The reason is that the predictions of the Rayleigh instability, for a flow regime with coherent structures, are valid only at the special value of $\Lambda_f/\delta_\omega$  achieved in the universal self-preserving state. 

\end{abstract}

\maketitle

\section{Introduction}
Beginning from Kelvin, Stokes and Rayleigh (see Darrigol\cite{darrigol2005worlds}) in the nineteenth century, instability has been conjectured to be an explanation for why flows of real, slightly viscous fluids behave so differently from known steady solutions of Euler’s equations.  Small perturbations have been shown to destabilize a free shear layer by different approaches - matching pressures below and above a vortex sheet (by Kelvin), analysis based on migration of the vorticity in a vortex sheet due to the induced velocity obtained from the Biot-Savart relation (by Helmholtz), and calculation by matching pressures between the outer irrotational flow and inner rotational flow for a piece-wise linear velocity profile (by Rayleigh, who also derived a generalized equation cast as an eigen value problem). 
While these results and later developments are useful in describing the early evolution of perturbations in free shear layers, the central question remains whether one can identify the connection between the initial instability, which is strongly initial-condition dependent, and the asymptotic self-preservation state observed in experiments (e.g. Brown \& Roshko\cite{brown1974density}). Though turbulence is generally considered to be a strongly non-linear phenomenon, an understanding from an instability point of view has been sought beginning with Malkus \cite{malkus1956outline} for a channel (which ultimately proved to be unsuccessful).  Such attempts have been more useful in the case of the free-shear layer (see Roshko\cite{roshko2000problem}). There have been attempts to use linear (e.g. Gaster \emph{et al.}\cite{gaster1985large}) or weakly non-linear (e.g. Monkewitz\cite{monkewitz1988subharmonic}) stability theories to describe the growth of perturbations in a turbulent free shear layer. Though such studies have achieved some success in predicting growth rates of externally imposed perturbations (see Ho \& Huerre\cite{ho1984perturbed} for a review), they have been controversial \cite{hussain1983coherent,husain1995experiments}. The most significant development in connecting instability theory with asymptotic behavior has been the model of Morris \emph{et al.} \cite{morris1990turbulent} . In this model, the self-preservation growth of a shear layer is computed by balancing the energy gained in the instability modes (as predicted by linear stability theory) with the loss of mean flow energy. 

Before proceeding further with the discussion on instability, we note that two controversies exist in understanding the asymptotic evolution of the fully three-dimensional flow in planar free shear layers. First is the role of three-dimensionality \cite{d2013organized,suryanarayanan2017insights,brown2012turbulent} in determining the large scale evolution (spread rate and growth mechanism) before and after the `mixing transition' \cite{konrad1977experimental,dimotakis2000mixing}.  The second is whether the asymptotic self-preserving state is universal or dependent on initial conditions \cite{oster1982forced,narasimha1990utility,brown2012turbulent,george2004role}.  Extensive simulations (using the vortex-gas model, to be described in detail in Sec. 2) of a 2D temporally evolving free shear layer by Suryanarayanan, Narasimha and Hari Dass \cite{snh} (SNH henceforth) have provided new insights. In this work, it was shown that a temporally evolving free shear layer in a gas of point vortices, in a streamwise periodic domain $L$, exhibits three distinct regimes. The final Regime (RIII)is a long term solution involving a single structure in $L$ (found to be related to equilibrium statistical mechanics\cite{joyce1973negative}) and the early evolution (RI) is strongly dependent on initial conditions. The intermediate regime (RII) was, however, found to be a self-preserving state that was independent of initial conditions and domain size. Remarkably, it exhibited universality in growth rate and other flow parameters. A Galilean transformation of this spreading rate from SNH is in broad agreement with the self-preservation spread rates in high Reynolds number experiments (on spatial mixing layers whose ratio of low to high speed velocity is greater than one third, e.g. refs.\cite{spencer1971statistical,d2013organized}).This comparison is summarized in Suryanarayanan \& Narasimha \cite{suryanarayanan2017insights}, who also demonstrate existence of universal self-preserving states in spatial vortex-gas shear layers). One possible interpretation of this agreement is that the large scale evolution of even post-mixing-transition plane free shear layers is dominated by quasi-2D coherent structure interactions and thus could be described, in part, by purely 2D mechanisms.(Such an interpretation, broadly dependent on a multi-pole expansion of the vorticity field, would suggest that the universal self-preservation observed in the 2D vortex gas is broadly applicable in real shear layers.).Regardless of the relevance to real shear layers, however, the universality observed for the 2D free shear layer raises the following question in the present context: \emph{how does a deterministic approach such as instability analysis (which is dependent on initial conditions) eventually yield the observed self-preservation state, which is independent of initial conditions, for  the 2D inviscid case?}  This context forms the basis of this study.

The present paper attempts to understand the universality observed by SNH in the context of instability. The same vortex-gas method used by SNH is employed. This method does not seek a differential equation for a velocity perturbation but rather a velocity field based on the vorticity field at any instant of time, through the Biot-Savart relationship. The vorticity in the next time step is moved in accord with that velocity field; that is, the vorticity of a fluid particle remains constant as it moves in the velocity field.  An important aspect of this form of the dynamics is that it permits nonlinear evolution for all times; it does not have the limitation of linear stability theory which allows a small perturbation in the vorticity field to be transported only with the mean velocity in the streamwise direction. Furthermore, the vortex-gas method provides a simple way to ‘understand’ the mechanics.  That is, even an accurate solution of the Euler equations may be difficult to ‘understand’ from the solution for the velocity field because the pressure field (which could be obtained) is complicated, and its gradient (which  is the force that acts to change the velocity field locally) would not be easy to interpret in the unsteady flow.  For a Biot-Savart approach, which is readily visualized, the dynamics is in the conservation of vorticity following a particle and in the velocity field which is necessarily connected with the vorticity through the Biot-Savart relationship.

The paper is organized as follows.  In Sec. \ref{sec2}, the governing equations and numerical method are described. In Sec. \ref{sec3}, we study linear instability using the vortex-gas method and compare the solution with the result from Rayleigh theory. Section \ref{sec4} focuses on the emergence of and the role played by coherent structures. In this context, the usefulness of the linear instability theory of Rayleigh is evaluated on the one hand and a monopole approximation as in the linear Karman-Rubach-Lamb (KRL) instability on the other. In Sec. \ref{sec5}, we focus on the applicability of instability approaches, including the Morris model, and a proposal for understanding the universal self-preservation state found by SNH and in the present simulations. We conclude in Section \ref{sec6}.  Appendix A compares and contrasts the eigen modes of Rayleigh theory with results computed from the vortex-gas calculations. A note on the Krasny type desingularization \cite{krasny1986desingularization} is presented in Appendix B. Appendix C describes the result from a case with initial conditions for a thick shear layer, which covers the entire phase of relaxation to self-preservation and supports the observations and conclusions made in Sections \ref{sec3}-\ref{sec5}.Appendix D provides details of the Morris model as applied to incompressible 2D temporal free shear layers.

\section{Computational Setup}
\label{sec2}
The dynamics of the 2D temporally evolving shear layer are perhaps most directly understood by studying the evolution of the 2D vorticity equation, obtained by taking the curl of the Euler equations (which eliminates pressure) and defining vorticity $\omega = \partial v / \partial x - \partial u / \partial y$. For the inviscid plane flow considered here, such a vorticity equation is conceptually very simple, namely

\begin{equation}
\frac{D\omega}{Dt} = \frac{\partial \omega}{\partial t} + u \frac{\partial \omega}{\partial x} + v \frac{\partial \omega}{\partial y} = 0
\label{vorticyeqn}
\end{equation}

The velocity is an integral of the vorticity through the Biot-Savart relationship, i.e.  $\mathbf{u} = \textrm{curl}^{-1} \mathbf{\omega}$.  In 2D,  $u = \partial \psi / \partial y, v=-\partial \psi / \partial x$ with the stream function, $\psi(\mathbf{r})=-(1/2 \pi) \int \omega( \mathbf{r'}) ln|\mathbf{r-r'}|  \mathbf{dr'}$.

Of course today there are numerical methods that could be used to solve for the evolution of a continuous distribution of vorticity (see ref.\cite{cottet2000vortex} for a review).  There is attractiveness, however, in an initial discretization of the vorticity field into small areas in which the vorticity is then concentrated in a point vortex that has the same circulation as the vorticity in the small area which it represents. Such a discretization allows a very simple description and comprehension of the mechanics in the evolution of a two-dimensional shear flow as the strength of each vortex remains constant and the integral of the vorticity field, from which the velocity is obtained, is reduced to a simple algorithm for summing the contributions of each point vortex (including an infinite array based on an initial domain repeated to infinity).  An additional advantage is that the numerical accuracy of solutions for very long times can in part be conveniently assessed through any change over time in the Hamiltonian for this system. While this method appears simple, there are mathematical theorems (eg. Beale \& Majda\cite{beale1982vortex}, Marchioro \& Pulvirenti\cite{marchioro1993}) that prove the convergence of the vortex-gas to weak solutions of the Euler equations under appropriate limits (see refs.\cite{snh,suryanarayanan2017insights} for further discussion). The use of the point-vortex approximation has a long history which includes the early work of Rosenhead\cite{rosenhead} and subsequently of Hama and Burke\cite{hama1960rolling}, Acton\cite{acton}, Delcourt \& Brown\cite{delcourtbrown}, Aref \& Siggia\cite{arefsiggia} and others. Most recently, extensive simulations (using up to 32,000 vortices) of the temporal vortex gas shear layer by SNH\cite{snh} yielded fresh insights on 2D free shear layers, particularly the role of initial conditions, the universality and self-preservation of an intermediate regime and relation of the final state to equilibrium statistical mechanics.

\begin{figure}
	\centerline{\includegraphics[width=5.0in]{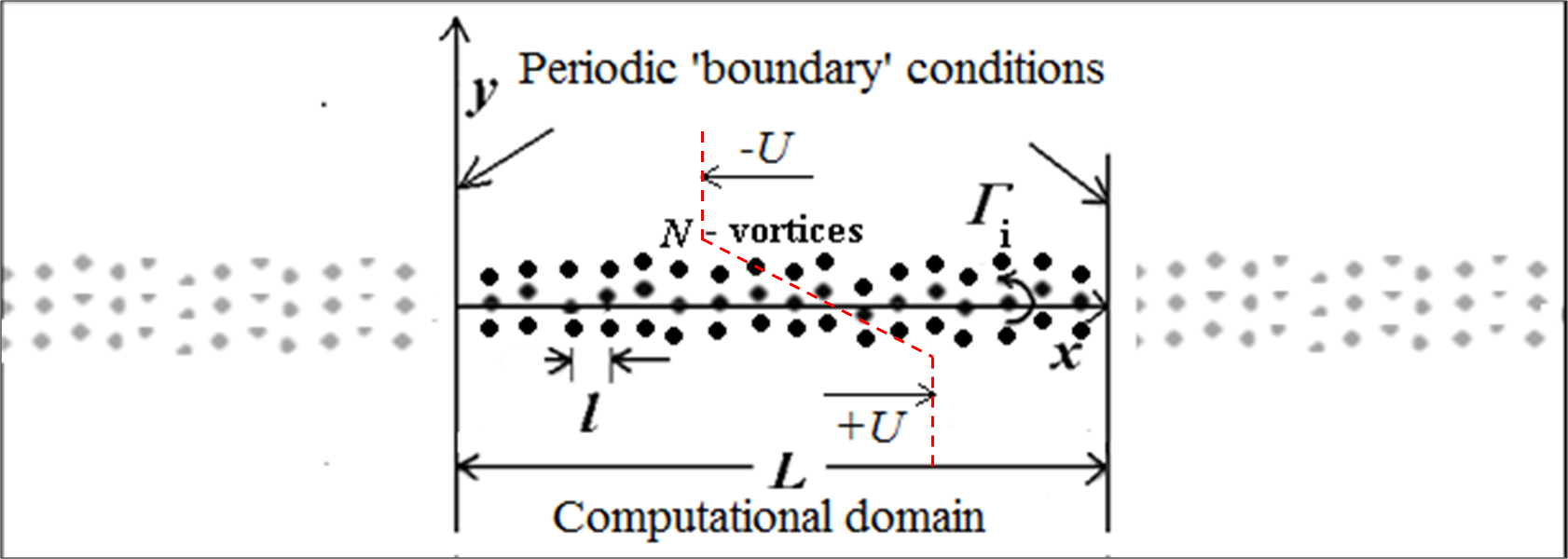}}
	\caption{\label{Fig1} Details of the present setup showing the initial vortex locations.  The corresponding piece-wise linear induced velocity profile is shown in (red) dashed lines.}
\end{figure}

In the present work, the same approach and numerical method are used as in SNH; i.e. a singly-periodic Hamiltonian system of point vortices is numerically integrated using double precision RK-4.  The only differences between SNH and the present study are the class of initial conditions explored, the temporal regimes of focus and the methods of analysis. The present approach involves representing a continuous vorticity field using an array of $N$ point vortices that are by default equi-spaced in $x$ and $y$ (with spacing $l$) at $t=0$. Note that in contrast to SNH and most earlier approaches, we adopt initial conditions that have many `layers' or rows of point vortices and discretize a small area of vorticity rather than a single line of point vortices at $t=0$. 

The present arrangement (as in Fig.\ref{Fig1}) has $N$ point vortices in a periodic domain of length $L$ containing an inviscid fluid. As a consequence of the periodicity, for each vortex in the domain that exists at $(x,y)$, there exist vortices at $\{(x+kL ,y)\} ; k=\{-\infty…-2,-1,0,1,2,...\infty\}$. The velocity field in the domain at any time $t$ is completely determined by the instantaneous location of all the vortices, which includes the vortices in all other periodic domains. This summation of the Biot-Savart relation over all the periods leads to the closed form expression (see Lamb\cite{lamb1932hydrodynamics} for details),

\begin{equation}
u(x,y) = \sum_{j=1,j \neq i}^{N} \frac{ - \gamma_j}{2L} \frac{\sinh(2\pi(y-y_j)/L)}{\cosh(2\pi(y-y_j)/L)-\cos(2\pi(x-x_j)/L)}
\label{equ}
\end{equation}

\begin{equation}
v(x,y) = \sum_{j=1,j \neq i}^{N} \frac{\gamma_j}{2L} \frac{\sin(2\pi(x-x_j)/L)}{\cosh(2\pi(y-y_j)/L)-\cos(2\pi(x-x_j)/L)}
\label{eqv}
\end{equation}

where $x_j, y_j$ and $\gamma_j$ are respectively the $x$ and $y$ locations and the circulation of the $j^{\textrm{th}}$ vortex. The governing equation for vorticity in two dimensions (eqn.\ref{vorticyeqn}) results in the vorticity at any point being advected by the velocity at that point (which in turn is determined by the vorticity everywhere via the Biot-Savart relation). Therefore, for the system of point vortices, 

\begin{equation}
\frac{dx_i}{dt} = u(x_i,y_i)
\label{dxidt}
\end{equation}

\begin{equation}
\frac{dy_i}{dt} = v(x_i,y_i)
\label{dyidt}
\end{equation}

where the velocity is given by eqns.(\ref{equ},\ref{eqv})  Hence this leads to a system of $2N$ coupled first order ODEs that can be solved as an initial value problem. We set the circulation of each vortex to be equal to $\gamma = 2UL/N$, so that the $x-$velocity is $-U$ and $+U$ above and below the layer respectively (with velocity difference $\Delta U=2U$; see Fig.\ref{Fig1}). We adopt a time step of 0.1 times the ratio of the initial inter-vortex spacing to the velocity difference of the two far-field velocities.

One of the conserved quantities of this system is the Hamiltonian (often also called Kirchoff’s function) that is defined as\cite{delcourtbrown}

\begin{equation}
\mathcal{H} = \frac{-1}{8\pi} \sum_{i=1}^{N} \sum_{j=1,j \neq i}^{N} \gamma_i \gamma_j \ln\big[\frac{1}{2}(\cosh(2\pi(y_i-y_j)/L)-\cos(2\pi(x_i-x_j)/L))\big]
\label{Hamiltonian}
\end{equation}

The present numerical scheme and time step ensure that the Hamiltonian is preserved to better than 1 in $10^5$ of its initial value.   

A momentum thickness, $\theta=\int ((1/4) - (u/\Delta U)^2) dy$ , and a vorticity thickness $\delta_\omega = \Delta U/\textrm{max}[\partial u/ \partial y]$, are computed from the $x$-averaged velocity profile. The stream function $\psi$ is computed using the expression (obtained from the definitions of $u$, $v$ and $\psi$)

\begin{equation}
\psi(x,y) = \frac{-1}{4\pi} \sum_{j=1}^{N} \gamma_j \ln[\frac{1}{2}(\cosh(2\pi(y_i-y_j)/L)-\cos(2\pi(x_i-x_j)/L))] 
\label{psi}
\end{equation}

Desingularization (e.g. Krasny\cite{krasny1986desingularization}) has not been adopted in most of the simulations presented in this script as it does not affect the major conclusions. This is elaborated in Appendix B, which also presents some results and discussion on convergence. 

\section{Linear instability of a constant-vorticity shear layer}
\label{sec3}
To understand instability from a mechanistic point of view, we first consider a problem for which a closed form solution exists: the growth of small perturbations in a piece-wise linear velocity profile (i.e. constant vorticity strip of finite thickness). The application of the Rayleigh equation leads to an exponential growth for a range of wavenumbers ($2\pi \delta_\omega/\Lambda)$ less than $1.28$ with a maximum exponent of $0.402$ occurring at a wavenumber of $0.798$ (i.e. $\Lambda/\delta_\omega=7.87$).  The vorticity eigen function is a pair of vortex sheets, one located at the upper edge and the other at the lower edge of the initial constant vorticity profile, and each sheet has a strength that varies sinusoidally in $x$. 

\begin{figure}
	\centerline{\includegraphics[width=6.5in]{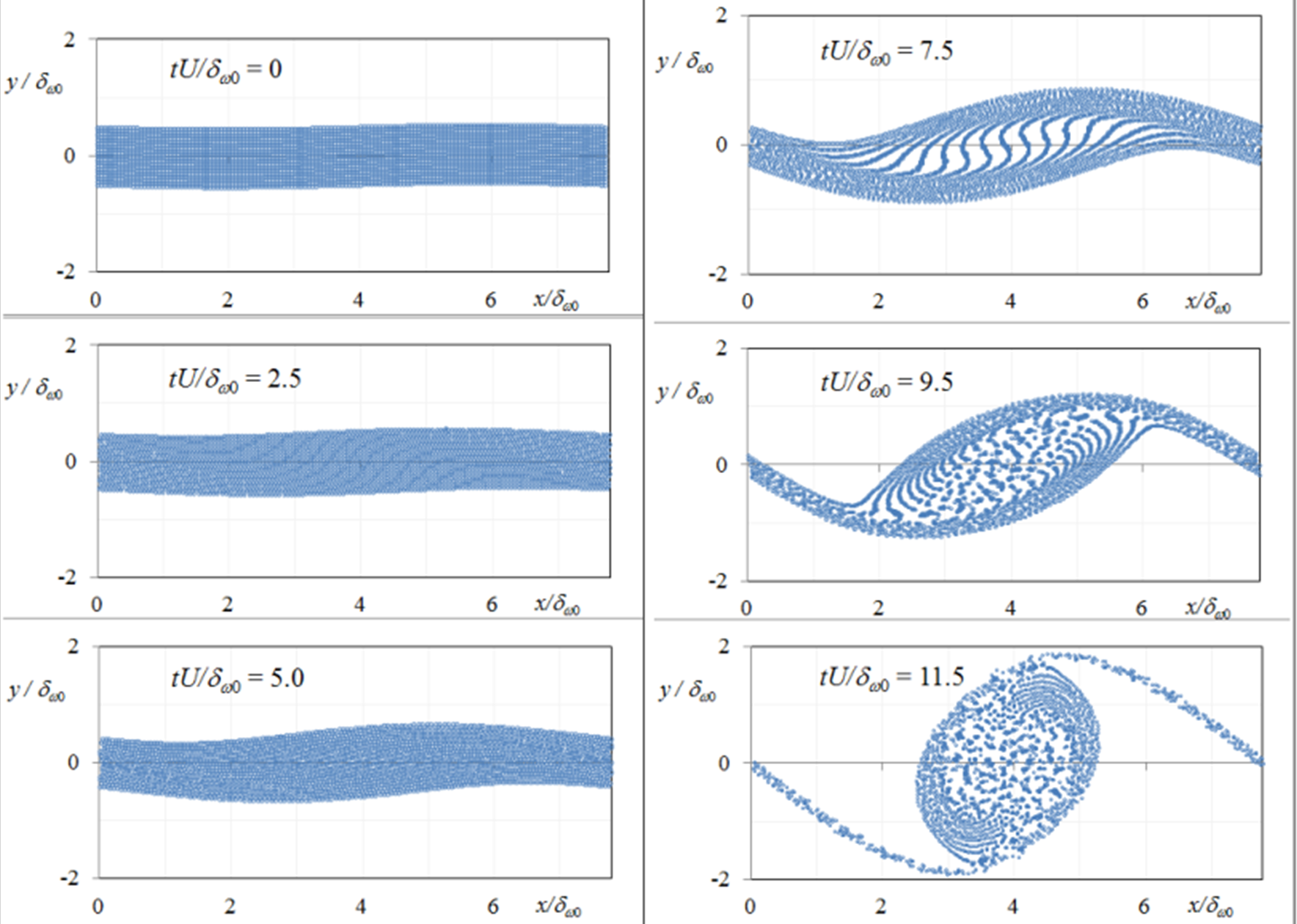}}
	\caption{\label{Fig2} Temporal evolution of the vortex gas representation of the constant-vorticity free shear layer with a single mode perturbation showing the migration of vorticity. }
\end{figure}

\begin{figure}
	\centerline{\includegraphics[width=6.5in]{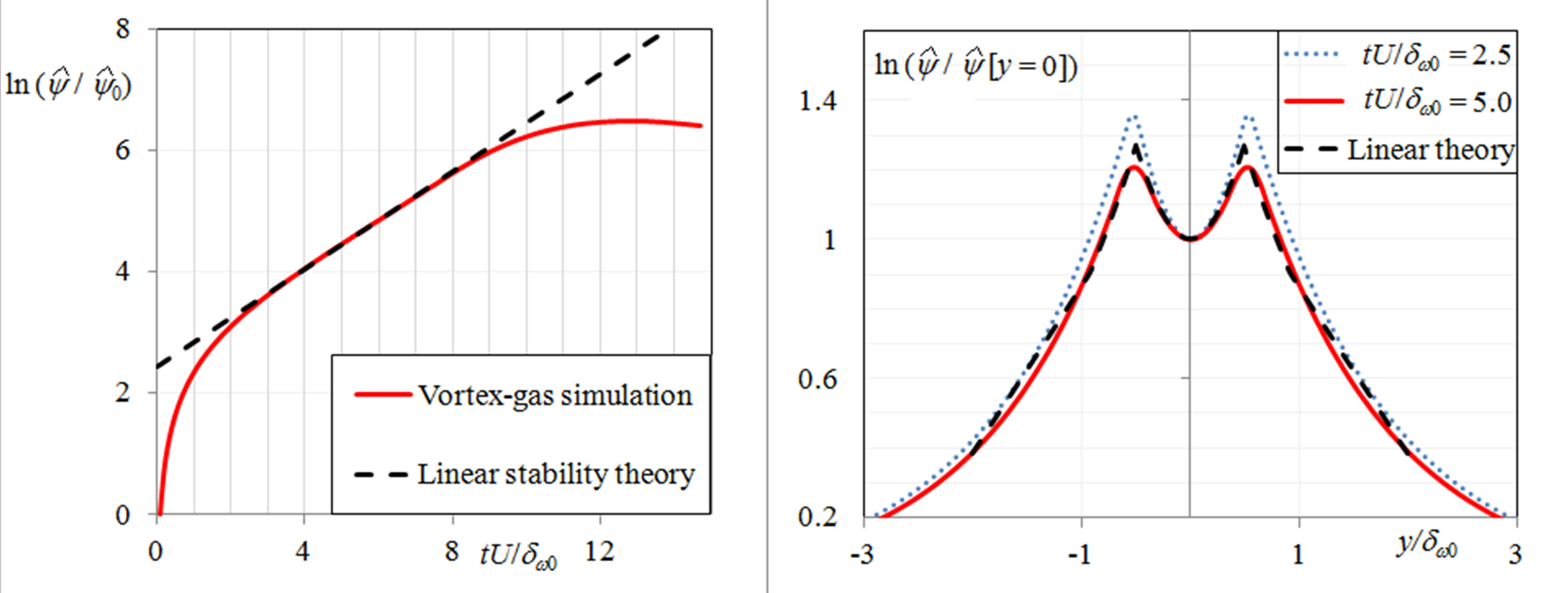}}
	\caption{\label{Fig3} \textbf{(A)} Evolution of the modal amplitude of the perturbation stream function at $y = 0$, shown in log-linear scale. Note that the growth is exponential (linear in log scale) for $tU/\delta_{\omega0} \sim 2.5$ to $8.5$  and the exponent is in excellent agreement with that predicted by Rayleigh theory. \textbf{(B)} The distribution of perturbation streamfunction from the full vortex-gas calculation is close to that created by the pair of vortex-sheets in Rayleigh theory. }
\end{figure}

The most amplified mode in the vortex-gas is simulated first, using 19 rows of vortices with 140 vortices per row ($N = 2660$), equi-spaced in both $x$ and $y$ (contributing to an unperturbed `thickness' $\delta_{\omega 0}=18l$). At $t = 0$, all the vortices are given a sinusoidal perturbation in $y$, with a wavelength $\Lambda_f$ exactly equal to the length $L$ of the periodic domain and an amplitude of $0.005\Lambda_f$ ($\Lambda_f/\delta_\omega=140/18 \approx 7.8$ ;  the subscript $f$ stands for fundamental). It can be seen from Fig. \ref{Fig2} that the initial perturbation grows in time, and is marked by a migration of vorticity. Unlike the Rayleigh approximation no vortex sheets are introduced; instead the shape of the field of vorticity changes. The evolution of the perturbation eventually leads to a roll up into a ‘coherent’ vortical structure. Figure \ref{Fig3}A shows that after a short initial transient of $tU/\delta_{\omega 0} \sim 2.5$ (the transient can be eliminated by initializing with a `Biot-Savart eigen mode' initial condition discussed in Appendix A), the amplitude of the perturbation (stream-function) grows exponentially. The exponent closely agrees with the Rayleigh solution, which provides a demonstration of the theorem that the point vortex approximation provides a weak solution of the Euler equations. Additional notes on convergence with $N$ can be found in Appendix B. The exponential growth begins to saturate at $tU/\delta_{\omega 0} \sim 8.5$. Note that the time of this departure approximately coincides with the formation of the coherent structure.  Figure \ref{Fig3}B shows the distribution of the perturbation streamfunction in $y$ observed in the vortex-gas solution and compares it with that produced by the two vortex-sheets, one at each edge of the constant vorticity layer constituting the eigen function of Rayleigh. Again there is close agreement.

Using de-singularized vortices gives similar results, but the emergence of chaos within the structure is delayed.  As long as the desingularization radius is small compared to the wavelength of the instability, the evolution of the coarse grained vorticity field and the conclusions henceforth are largely unaffected. More details are presented in appendix B.
The results for single mode analysis shown in Figs. \ref{Fig2} and \ref{Fig3} compare and contrast the Rayleigh and Kelvin-Biot-Savart approaches to instability, ultimately demonstrating the equivalence of the two approaches for the linear evolution (Regime Ia) despite the different nature of the approximations (Rayleigh being a linear approximation to the strong solution to the Euler equation while the vortex-gas provides a weak solution). The Kelvin-Biot-Savart approach provides a vorticity-transport-understanding of the Rayleigh eigen-function. The remarkable and important fact is that for inviscid flow all of the dynamics (linear stability, saturation and non-linearity) is contained in the simple Kelvin-Biot-Savart mechanics.   The instability of a finite thickness and uniform distribution of spanwise vorticity results in the migration of point vortices in different regions, and hence also of the vorticity which remains constant following a particle. This leads to regions of accumulation and depletion of this vorticity.  In the Rayleigh instability theory this is approximated by two non-physical vortex sheets placed at $\pm y=\delta_{\omega 0}/2$ having sinusoidal variations in strength.  With hindsight this approximation to the actual migration of the vorticity is self-evident; it is also clear why the same approximation is accurate in the linear region but fails, of course, to account for the saturation of the amplitude due to the non-linearity.

\section{Emergence and role of coherent structures}
\label{sec4}

\begin{figure}
	\centerline{\includegraphics[width=6.5in]{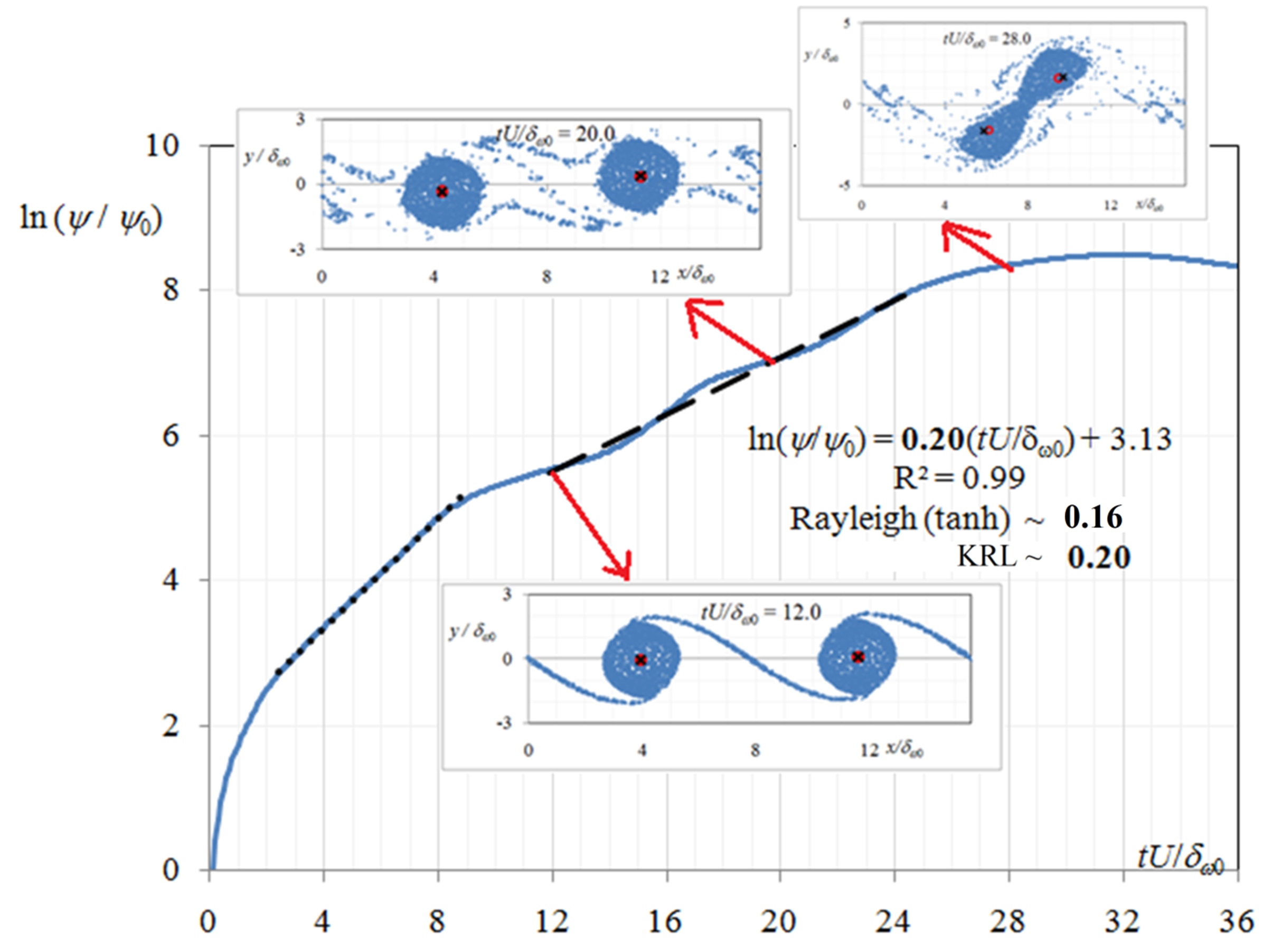}}
	\caption{\label{Fig4} Evolution of the (logarithm of) perturbation stream function amplitudes corresponding to the subharmomic. Following the `saturation' of the initial exponential growth, the subharmonic enters a new regime, also growing exponentially over $12\le tU/\delta_{\omega 0} \le 24$ with an exponent of $0.20$ as shown by a dashed line, which is closer to the KRL prediction than that of Rayleigh. Also shown are snapshots, which show the evolution of the structures in the full simulations at $tU/\delta_{\omega 0} = 12$, $20$ and $28$, superimposed with the evolution of monopoles (shown by crosses).  The monopole approximation (as well as the KRL instability) is useful from the time coherent structures first form to the time they are about to merge. }
\end{figure}

The simulation presented in Section 3 showed the linear and nonlinear evolution of a single mode culminating in a roll up into a coherent structure. Due to the choice of $\Lambda_f =L$, the single coherent structure does not have the opportunity for merging with other structures as wavelengths greater than $L$ are not permitted.  In a simulation with much larger domains (as adopted in SNH), the subsequent evolution may be expected to be governed by the interaction of several structures. As a first step towards understanding such a process, we simulate here the evolution of the free shear layer with two modes, essentially doubling $L$ from the previous section and imposing an additional longer wavelength disturbance in the initial condition.  The resulting simulation has 5320 vortices, arranged in 19 rows as before. At $t = 0$, the $y-$positions of the vortices are perturbed with two modes. The first mode has a wavelength $\Lambda_f=L/2$ and an amplitude $a_f = 0.005 \Lambda_f$, and we shall call it the fundamental (as in the fluid-dynamical literature).  The second mode, called the subharmonic, has a wavelength $\Lambda_s=L=2\Lambda_f$ and amplitude $a_s=0.001 \Lambda_f$. As in the case of the single-mode perturbation, the waves are found to roll up into structures at $tU/\delta_{\omega 0} \sim 8$. Beyond this time, the evolution seems to be dominated by what appears to be an interaction of the two structures, as seen in Fig. \ref{Fig4}. 

\begin{figure}
	\centerline{\includegraphics[width=4.0in]{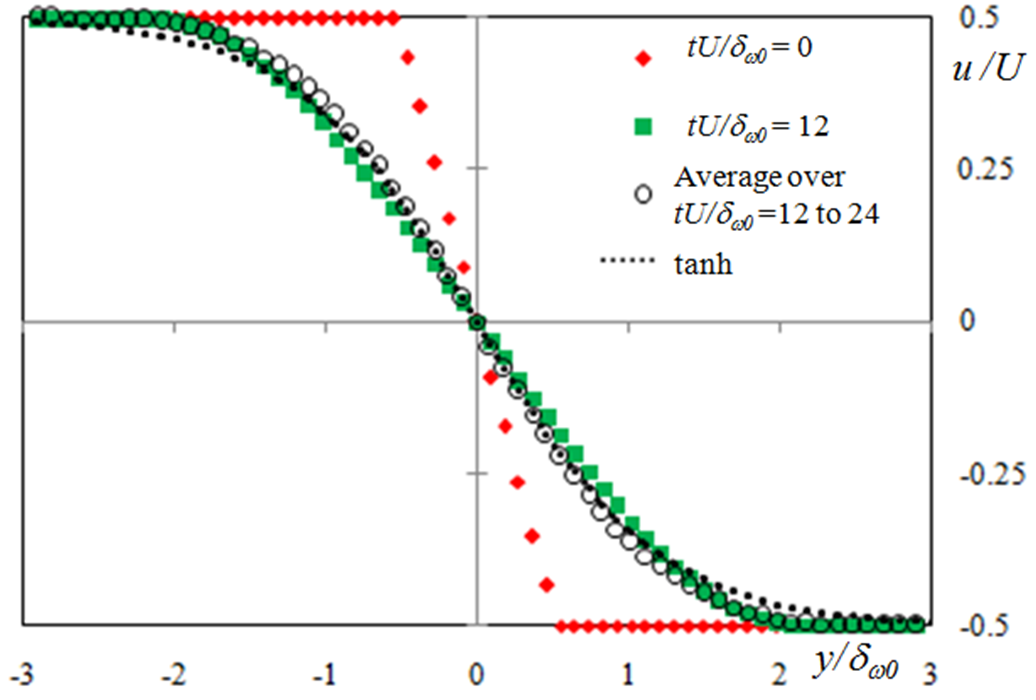}}
	\caption{\label{Fig4b} Evolution of the velocity profile with time.  The $x-$averaged velocity field relaxes from an initially piece-wise linear profile to one that is well approximated by a tanh profile beyond $tU/\delta_{\omega 0} \sim 12$ }
\end{figure}

The evolution of the subharmonic amplitude is plotted in Fig. \ref{Fig4}.  Both the fundamental (not shown) and the subharmonic initially grow exponentially until the saturation of the fundamental.  After this, the subharmonic enters a new regime of exponential growth.   Interestingly, the observed exponent (0.2) is larger than the one predicted by the application of the Rayleigh theory to a local new base flow (which has grown in thickness following the roll-up, and is closely approximated by a tanh velocity profile, see Fig. \ref{Fig4b}). The Rayleigh exponent for the tanh profile with thickness equal to that observed at  $tU/\delta_{\omega 0} = 12$ is 0.12.  If the vorticity thickness of the velocity profile time averaged between  $tU/\delta_{\omega 0}$ of 12 and 24, the duration over which exponential growth is observed, Rayleigh theory predicts an exponent of 0.16.

To understand this lack of agreement, a closer look is taken at the evolution of the layer during this time.  A natural, simplified way to describe the flow field at  $tU/\delta_{\omega 0} =12$ is to consider each structure as a single point vortex (referred to as a monopole), located at the respective centroid of the structure and with a strength equal to the circulation of the structure (which, for the vortex gas case, is equal to the sum of the strengths of all the point vortices contained within the structure), rather than to consider a uniform-in-$x$ base flow with perturbations (as in Rayleigh theory). Since the number of vortices located in the thin braids is small (less than 10\%) compared to the number in the structure, we can crudely approximate the locations of the monopoles based on the centroid of vortices with $x<L/2$ and $x>L/2$ respectively. Such a monopole approximation initialized at the locations of the centroids is shown in the inset at  $tU/\delta_{\omega 0} =12$.  The system of two monopoles (shown as black crosses) is now simulated and it is found that the monopoles closely follow the trajectories of the centroids (shown as red circles) and this approximation seems sufficient to provide a simple description of the evolution, before the onset of merger of the two structures ( $tU/\delta_{\omega 0} >28$ approximately). 

A theory of growth of small perturbations to a system of such monopoles was first proposed by Karman \& Rubach\cite{v1912mechanismus} and reproduced in detail by Lamb\cite{lamb1932hydrodynamics}, and it shall therefore be referred to as ‘Karman-Rubach-Lamb’ (KRL) theory. The basis of this instability is that any perturbation to either the strengths or the locations of the monopoles in this array is unstable because the velocity of at least one vortex due to all of the vortices to the left will not be exactly equal and opposite to the velocity induced by all the vortices to the right of it.  That vortex will therefore move, and the resulting motion increases the imbalance. In the limit of the monopole displacements from their ‘base’ locations ($x=L/2 \pm L/4,y=0 )$ being small compared to the distance between them $(\Delta x_m = L/2)$, the evolution of their perturbations can be analytically calculated to be exponential, with an exponent of $\Gamma \pi / (4(\Delta x_m )^2)$ (see Lamb\cite{lamb1932hydrodynamics}). For the present flow with  $12<tU/\delta_{\omega 0}<24$, the displacement of the monopole locations is less than 15\% of the initial distance between them, and hence a ‘small displacement’ approximation seems reasonable. Indeed, the exponent computed using $\Gamma \pi / (4(\Delta x_m )^2)$, with $\Gamma=\gamma N/2$ and $\Delta x_m=L/2$, is $\pi U/L=0.20 U / \delta_{\omega 0}$   , is the same as the observed value (to within two significant digits) of the exponent in the present computations during the interval  $12<tU/\delta_{\omega 0}<24$,as shown in Fig. \ref{Fig4}.

Thus the evolution immediately following the emergence of coherent structures is governed by the monopole instability of KRL, and not by the instability of a base velocity profile (i.e. the Rayleigh approximation).

\section{The relaxation to universal self-preservation}
\label{sec5}
\subsection{Vortex-gas observations}

Sections \ref{sec3} and \ref{sec4} have shown the role of instability in determining the evolution of perturbations before and immediately after the formation of coherent structures. As discussed in Sec.\ref{sec4}, at the end of the KRL instability, the coherent structures merge resulting in the growth of the layer thickness. Experiments by Winant \& Browand \cite{winant1974vortex} carried out at moderate Reynolds numbers identified mergers as the mechanism of layer growth.  After several sets of mergers, the layer enters a universal self-preserving state \cite{snh}.

 The role of instability in the universal self-preservation state, is discussed in this section. We adopt the SNH terminology to identify the different regimes in the temporal evolution of the shear layer - RI refers to the ‘initial condition dependent’ evolution, which here is subdivided into I(a) and I(b) referring to the part of the relaxation before and after the emergence of coherent structures; RII refers to the self-preservation regime, found to be universal for the 2D vortex-gas by SNH; the finite domain driven RIII of SNH is not considered here.
  
A simulation of a constant vorticity shear layer initialized with broad band initial conditions is discussed in Appendix C. It was found to go through an initial, pre-coherent structure Regime I(a), where growth exponents of individual modes are accurately predicted by the Rayleigh model, consistent with findings in Section \ref{sec3}.  Regime I(a) is followed by the nonlinear roll-up and formation of coherent structures, constituting Regime (Ib). Here it was found that the KRL model (an application of which will be described shortly) predicts the observed growth exponents, but the Rayleigh model does not. Thus these observations establish the general applicability of the findings presented in Sec.\ref{sec4}. Subsequently the layer was found to evolve to the same self-preserving `reverse cascade' (Regime II) of SNH characterized by universal growth rate as well as spectral content. The detailed results of this simulation, that connects the results presented in earlier sections with Regime II, are presented in Appendix C. For the analysis of Regime II in the rest of this section, however, we shall use the data from SNH owing to the availability of larger sample size in terms of number of realizations; though broadly similar conclusions can be obtained from the thick sheet initial condition, see Appendix C, the reduced statistical scatter associated with the more extensive data enables more precise conclusions to be drawn. 

To analyze the role of instability, the spectrum of growth exponents is computed at a specified time from the spectral evolution of the perturbations, about the $x-$averaged base flow at that time, over a short duration ($20 l/U$). This is done using the following procedure. The (discrete) Fourier transform of the perturbation stream function at $y = 0$ is taken and then the logarithm of the ratio of the amplitudes for a given mode at two time instants (that are sufficiently close to ensure that the bulk statistics such as thickness have not changed significantly) is calculated to yield the local growth exponent. We compare the exponent distribution so obtained with the prediction of growth exponents based on the solution of the linearized Euler equation for the x-averaged base flow (approximated by an equivalent tanh profile) at that time. This shall henceforth be referred to as the `Rayleigh model'. 

\begin{figure}
	\centerline{\includegraphics[width=6.5in]{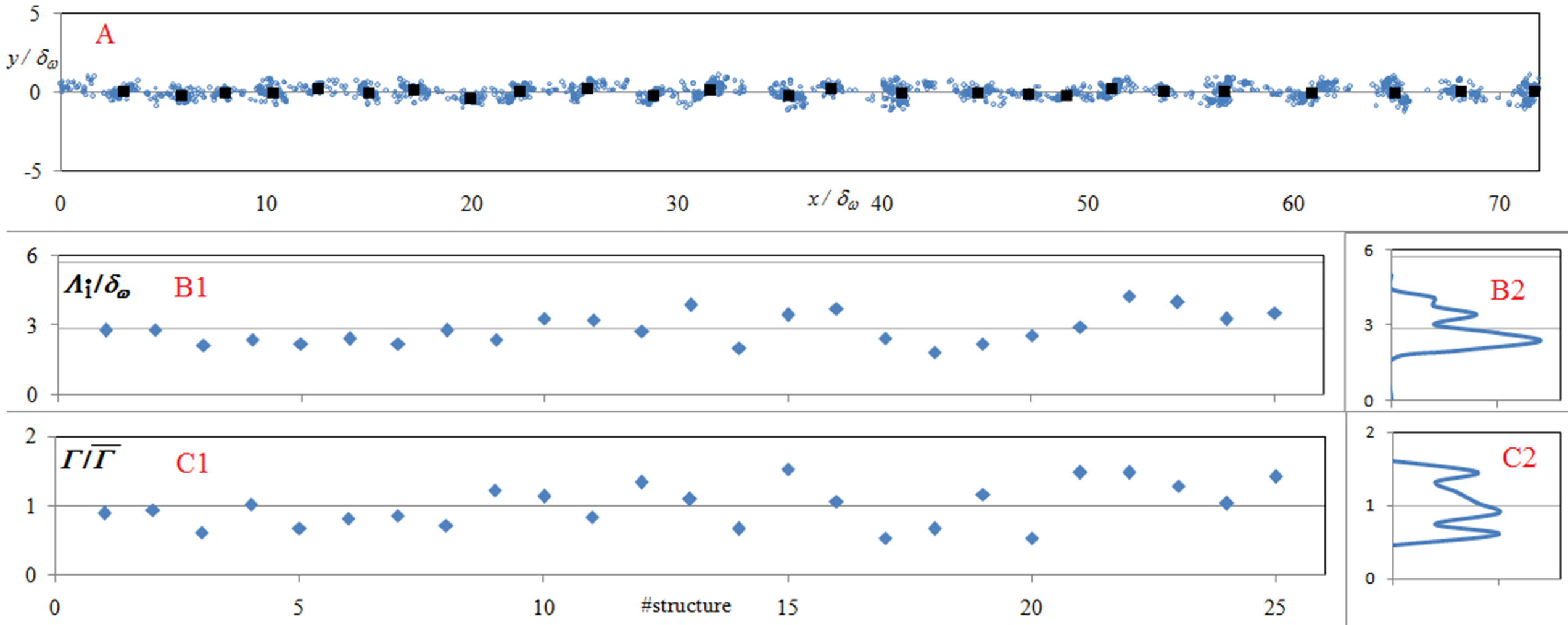}}
	\caption{\label{Fig5} \textbf{A.} Snapshot of the RII solution at $tU/l = 300$. The centroids of the $25$ identified structures are shown as black squares.  \textbf{B1.} Spacing $\Lambda_i  (=x_(i+1)-x_i)$ between the centroids of the structures, \textbf{B2.} PDF of this spacing; the $25$ data points used here is not sufficient for a reliable PDF, but suggests a possibility of a non-Gaussian distribution with a mean of $2.9 \delta_\omega$ (in broad agreement with literature) and a standard deviation of $24\%$.  \textbf{C1} and \textbf{C2},  normalized strengths ($\overline{\Gamma}=2.9 \delta_\omega \Delta U$)of the structures and the PDF (standard deviation is $31\%$ of mean) of the normalized strengths of the structures. Note that there is a correlation (but not a perfect match) between the spacing and strength. }
\end{figure}

SNH showed that a chaotic but structured motion develops in R II. The remarkable result of a universal growth rate for the mean vorticity thickness, independent of initial conditions, was obtained.  This universality is expected to extend to universal probability distributions for the spacing between structures, non-dimensionalized by the mean vorticity thickness, and for the strength of coherent structures, non-dimensionalized by the mean vorticity thickness and the mean free-stream velocity difference.  Rough estimates of such distribution functions, obtained from a small data set, are shown in Fig.\ref{Fig5} (insets B2 and C2).  The main result here demonstrates that in R II there is a system of structures, which do not have equal strengths and spacing.  Nevertheless, to simplify this system in order to compare with predictions of a linear `KRL model', the flow is approximated by a set of monopoles of equal strength and spacing corresponding to the  spectral peak in the perturbation streamfunction derived from the vortex-gas solution at that time. The KRL theory is used to calculate how such a system of monopoles would respond to a perturbation  of a given wavenumber, in say the $y-$displacements of the monopoles. 

\begin{figure}
	\centerline{\includegraphics[width=6.5in]{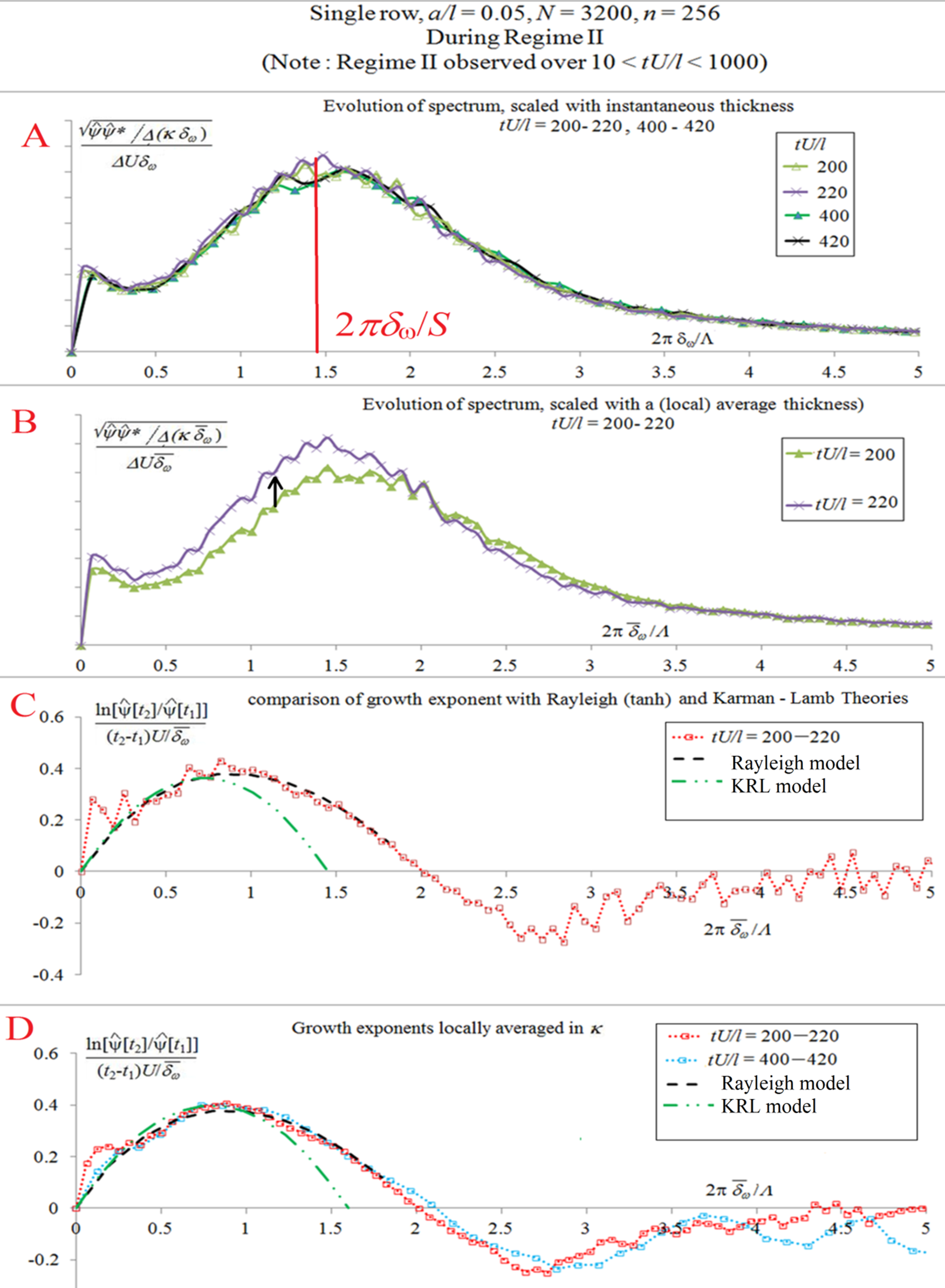}}
	\caption{\label{Fig6} Spectral evolution in Regime II computed for SNH case R1. Note that $\delta_\omega$ indicates vorticity thickness (computed from the $x-$averaged velocity field) at the specified time and $\overline{\delta_\omega}$ is the short-time averaged vorticity thickness (between t1 and t2 which are respectively 200 and 220 in B, C and for the red line in D, and 400 and 420 for the blue line in D). The arrow in B shows the shift in the spectral content with the small change in time.}
\end{figure}

The underlying instability mechanisms of this universal self-preserving regime are now  identified by computing the local growth exponent as a function of wavenumber ($\Lambda$) and comparing this exponent distribution with that predicted by the Rayleigh and KRL models.  It can be concluded from Fig.\ref{Fig6}A that this self-preservation flow has a continuous spectral content (in the limit), but it has a peak value. The wavelength (i.e. value of the abscissa) corresponding to this peak is denoted by $S$ ($(2\pi \delta_\omega)/S\sim1.6)$, and represents a measure of the most probable spacing between the coherent structures.

(Note that there is a difference between this value of $S=(2 \pi \delta_\omega)/1.6 \sim 3.9 \delta_\omega$ and the value $2.9 \delta_\omega$ obtained in Fig.\ref{Fig5}.  The former is obtained from the spectral distribution of the perturbation stream function amplitude at $y = 0$, whereas $2.9$ was obtained (from a relatively small sample) by subjectively identifying coherent structures and measuring the spacing between their centroids.   One reason for the difference is that when the perturbation has a continuous spectral content as opposed to being a single wavenumber, the peaks observed in the streamfunction field may be shifted to larger wavelength from that observed for the vorticity field (the spectrum of the latter is loosely the wavenumber squared times the former).  The physical interpretation of the shift is that structures that are in the process of merging may sometimes be identified as two separate structures from the vorticity field, but may effectively behave as a single monopole in terms of their long range influence. This effect is captured by the spectrum of the streamfunction field as it is an integral quantity.  Based on this argument, it was felt that when a continuous range of spacings is to be simplified into a system with a single spacing, the spacing obtained from the streamfunction field was a physically more convincing choice and hence adopted here as an input into the KRL model. If we had chosen 2.9 instead, it would have shifted the KRL predictions (green dashed lines) in Figs \ref{Fig6}C and D higher and towards the right, slightly worsening the agreement with the vortex gas data at small wavenumbers and slightly improving the agreement at long wavenumbers, and thus would not alter any major conclusion). 

The following observations can be made from Fig \ref{Fig6}D.  The growth exponent has a maximum value of around 0.4 at $(2\pi \delta_\omega)/\Lambda\sim 0.8$, which is about half of $(2\pi \delta_\omega)/S$ and hence represents the growth of the subharmonic of the spacing.  Surprisingly, this peak and the positive exponent distribution are accurately predicted by both the Rayleigh model and the KRL model (except at high-wavenumbers, which are beyond the monopole description).   That is, two opposite approximations, one involving small perturbations to a uniform-in-x base flow and the other which replaces each coherent structure with a single monopole, both appear to be in agreement with each other and provide a reasonable approximation of the flow in Regime II.  This is puzzling, because the predictions of the former depend only upon the thickness of the layer, while that of the latter depend only upon the dominant spacing of the structures. We shall return to explore this issue in Sec.\ref{sec5c}, but at present it is important to note the agreement with the Rayleigh model. If it were taken by itself, this finding would support the use of a Rayleigh inspired model to predict turbulent free shear layer evolution.  One such model has been proposed by Morris \emph{et al.}\cite{morris1990turbulent}.  

Fig.\ref{Fig6}B is a plot of the data in Fig.\ref{Fig6}A but scaled with an averaged thickness between $tU/l = 200$ and $220$. From this data the growth exponents are computed as a function of wavenumber and plotted in Fig.\ref{Fig6}C. A smoothed version is shown in Fig.\ref{Fig6}D, along with the results from analysis at $tU/l = 400$ to $420$. It can be seen that  this regime exhibits a self-similar universal function of not only the perturbation modes (Fig. \ref{Fig6}A)but also in their growth exponents with wavenumber (Fig. \ref{Fig6}D), when appropriately scaled with velocity difference and local layer thickness

\subsection{The Morris model prediction}

The two major assumptions made by the Morris model are: (i) the spread rate of the layer can be computed by balancing the energy gained by the perturbations in an integral sense and the energy lost by the growth of the layer, and (ii) the perturbation amplitudes (appropriately normalized to include the eigen function information) always grow exponentially in accordance with the application of the Rayleigh theory on the mean (in $x$) velocity profile at any given time.  The first assumption seems likely to be true for the vortex-gas shear layer. It is not clear at this point whether the second assumption would be valid, since Rayleigh theory was observed to provide accurate predictions in Regimes Ia and II but failed in the intermediate RIb. Furthermore, it is important to note that at any instant the spread of the layer determined by this model will depend on the existing spectral content, though the growth exponents themselves depend only on the thickness. This leads to the important question of whether there is an inbuilt mechanism in the model that could lead to universal self-preservation regardless of initial conditions.   

\begin{figure}
	\centerline{\includegraphics[width=5.0in]{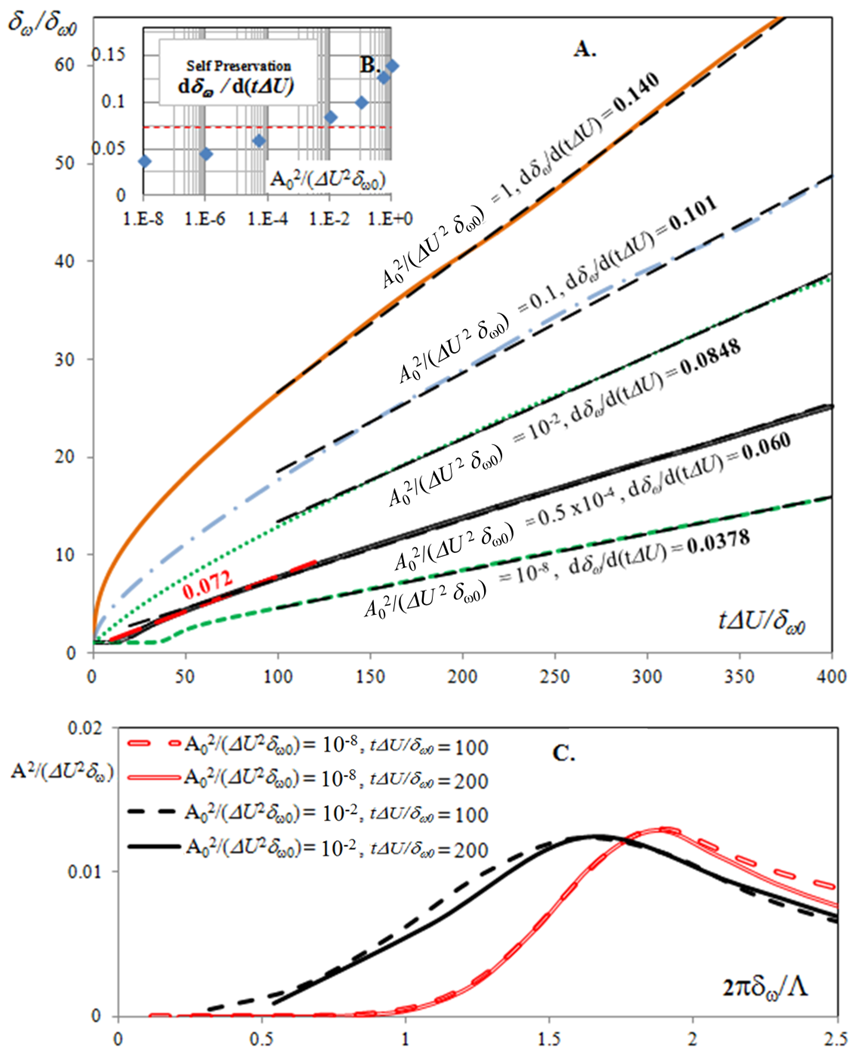}}
	\caption{\label{Fig7} \textbf{A.} Evolution of thickness in the Morris model for a temporal shear layer with broadband initial conditions of different amplitudes.\textbf{B.} Variation in the self-preservation spread rate with initial perturbation energy amplitude $A_0$. \textbf{C.}The self-preservation spectrum for different initial amplitudes. }
\end{figure}

In order to consider this question further, some preliminary simulations of the model based on Morris \emph{et al.}\cite{morris1990turbulent} are performed. This involves solving the coupled ODEs \ref{eqnD7} and \ref{eqnD8} (in Appendix D). These simulations were performed with a set of broadband initial conditions with each case having a different initial value of ${A_0}^2$, the kinetic energy per wavenumber (nondimensionalized by local thickness; see Appendix D for definitions). The results for the growth of the layer for the different initial conditions are shown in Fig.\ref{Fig7}A. It can been seen that while a state of constant spread rate (i.e. self-preservation) is reached in each of the cases (over nearly a logarithmic decade in time for some cases), the value of the spread rate itself is a function of $A_0$, which is associated with the initial conditions. Note that the self-preservation spread rate changes only very slowly with changes in $A_0$ (see Fig.\ref{Fig7}B) and this may have been a reason why the dependence was not noticed by Morris \textit{et al.}.  However, a difference in self-preservation spread-rate by a factor of nearly 4 is observed for the variation in $A_0$ considered in Fig.\ref{Fig7}. Further, the functional dependence of spread rate on $A_0$ also does not appear to settle down to any unique value.

This dependence of the self-preservation spread rate on the characteristic of the initial disturbance spectrum for a Rayleigh based model can be explained by the following.  While the growth exponents are independent of the initial energy amplitude $A_0$, it can be shown that the amplitude $A$ of any mode, at a time at which the amplitude of the most amplified mode reaches a specified value (say, at saturation), increases with increase in $A_0$. (The ratio of the amplitude of a given mode to that of the most amplified mode is $\textrm{exp}((G_\textrm{max}-G)t)$, and hence decreases with increase in $t$. For a given value of saturation amplitude, the time taken for the amplitude of the most amplified mode to attain that saturation amplitude is reduced for higher initial amplitudes.)  Hence, higher initial amplitudes lead to higher layer spread rates as more energy is distributed to the longer wavelengths that would subsequently become the most unstable modes under the increased layer thickness, and vice versa. This argument is consistent with the findings in Fig.\ref{Fig7}, particularly in Fig.\ref{Fig7}C, which shows the dependence of the final spectrum on the initial conditions. In each case, the spectrum evolves from the broadband initial condition to a self-similar distribution, which has a distinct maximum, but the width of the distribution and the precise location of the maximum depend on the initial condition. The Morris model therefore leads to a non-universal self-preservation. That is inconsistent with the SNH vortex-gas observations, which strongly support universal self-preservation (a unique, universal growth rate and self-similarity) over a wide-range of initial conditions for 2D free shear layers.

\subsection{Further comments on the failure of the Morris model and comments on the universal spacing to thickness ratio}
\label{sec5c}

The failure of the Morris model to reproduce the vortex-gas results on the universality of the growth rate merits closer consideration. This may seem somewhat surprising, since it is based on the Rayleigh model which accurately predicts the growth exponents of perturbation modes in RII.  Recall, however, the failure of the Rayleigh model in the intermediate Regime Ib.  The Morris model does not provide any special treatment to address this regime.

Consider therefore RIb and RII in greater detail. 	Why did the Rayleigh model fail in Regime Ib but provide good predictions in Regime II? One plausible reason why the Rayleigh approximation appeared to provide good predictions in Regime II is the following. After several mergers, the coherent structures are less orderly than what they were initially (RIb), and hence the perturbations they cause over the uniform-in-$x$ base flow are small relative to those in RIb and perhaps sufficiently small for the linear approximation of Rayleigh to hold. To investigate whether this argument is correct, let us closely examine the predictions of the Rayleigh model and KRL in RIb and RII.

\begin{figure}
	\centerline{\includegraphics[width=5.0in]{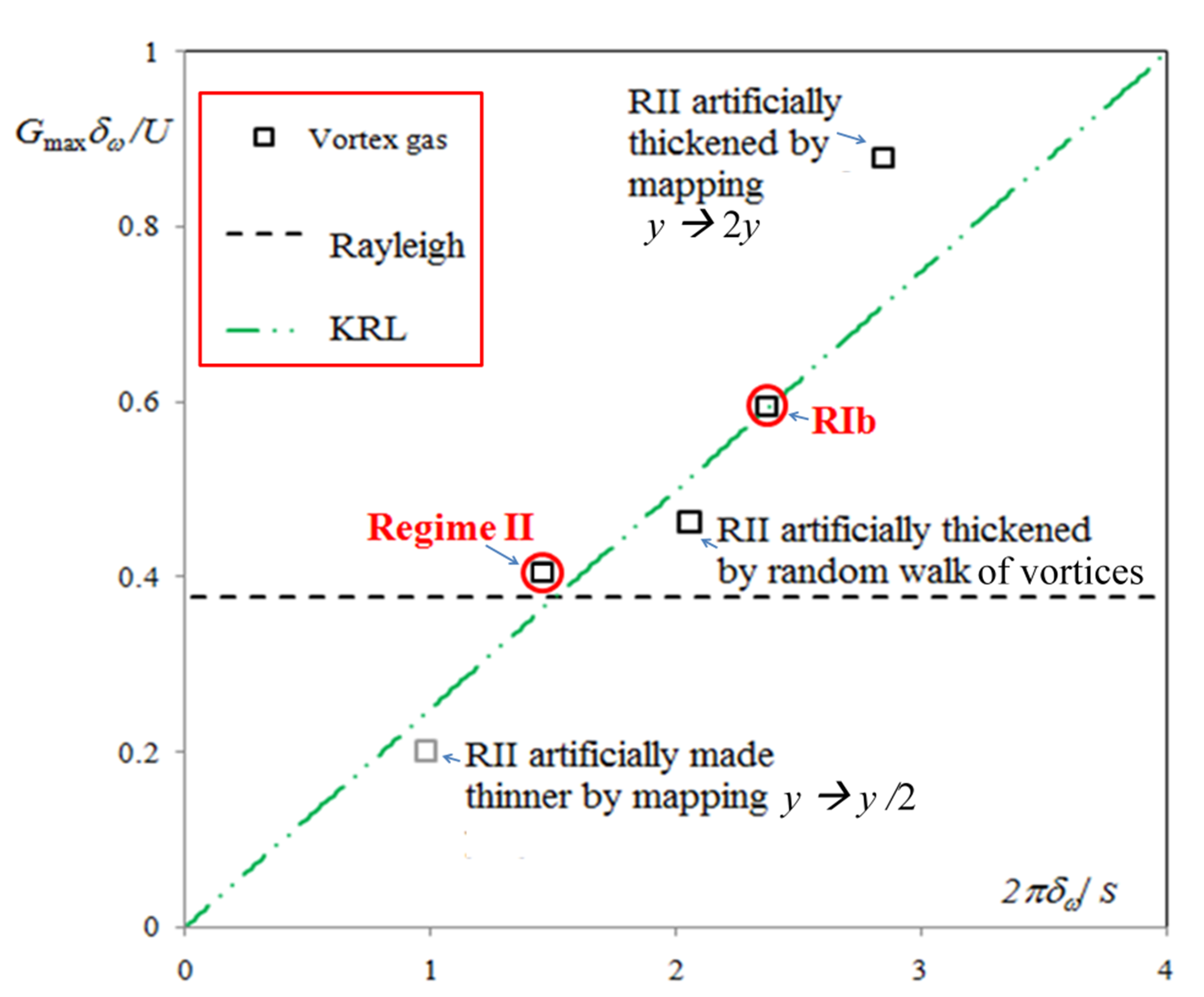}}
	\caption{\label{Fig8} Measured and predicted maxima (over wavenumber) of growth exponents scaled with thickness, plotted as a function of thickness-to-spacing ratio. }
\end{figure}

Note, however, that the Rayleigh model predictions are broadly based on the layer thickness and the KRL model predictions are broadly based on the spacing between structures. Figure \ref{Fig8} shows the maximum growth exponent, scaled with thickness, predicted by the Rayleigh model (black dashed line) and KRL (green dash-dot line) models for a given layer-thickness to structure-spacing ratio. As expected, since the Rayleigh model does not take structure information into account, its prediction does not depend on the spacing to thickness ratio and hence $G_\textrm{max}\delta_\omega/U =$ constant.  On the other hand, the KRL gives $G_\textrm{max}\delta_\omega/U = \textrm{constant} \times \delta_\omega/S$, as $G_\textrm{max} \propto 1/S$ in this model.  The vortex-gas data from regimes RIb (from the initially thick shear layer discussed in Appendix C) and RII (case R1, Fig. \ref{Fig6}), are both plotted on the Figure. 

The data can now be seen in new light. First,  RIb and RII both lie on the KRL line. Secondly, RII lies at the intersection of the KRL and Rayleigh models.  One set of interpretations of these observations is that (a) KRL is valid for all spacing-to-thickness ratios, and (b) the spacing-to-thickness ratio of RII is determined by the intersection of the KRL and Rayleigh models. 

To add more points to this plot, a set of calculations have been performed using three artificially generated configurations.  The first two are obtained by re-scaling the $y$ positions of all the point vortices. The first case involves scaling down y-location of all vortices by a factor of two, while the second case involves scaling them up by a factor of two. We also consider a third case in which the thickness of the layer is increased by approximately a factor of two, by a sudden ‘diffusion’. This is achieved by the addition of random numbers (drawn from a Gaussian distribution) to the $x$ and $y$ locations of each vortex (see Chorin\cite{chorin1973numerical} for details on simulating viscosity by random walk of point vortices).  It is important to note that no particular conservation law has been violated; these cases provide altered initial conditions for entirely new vortex-gas simulations. The purpose is to obtain configurations that retain the broad `chaotic but structured' character seen in Regime II but with different values of $2\pi \delta_\omega/S$.  The local growth exponents for these specified states are computed from the vortex-gas calculations, and the data from these simulations (after neglecting the anomalous very low wavenumber peak for the $2\pi \delta_\omega/S \approx 1$ point) are also plotted in Fig. \ref{Fig8}. It can be seen that the points roughly line up with the KRL predictions, but only for the Regime II value of $2\pi \delta_\omega/S$ are the predictions from the Rayleigh model acceptable. One might therefore conclude that the Rayleigh model is not uniformly valid in an evolution with coherent structures for arbitrary structure spacings. 
 
There is then further clarity on why the Morris model does not predict the universal self-preservation attained in the full (vortex-gas) simulation. The initial amplification for small perturbations is indeed in accordance with the modal growth based on the Rayleigh model. This leads to the formation of coherent structures whose spacing to the initial thickness ratio (and the relative amplitudes of the subharmonics and the shape of the spectrum) will depend on the initial conditions, and may or may not be the special value eventually attained in Regime II: only at this special value does the Rayleigh model provide good predictions of growth exponents.The growth exponents of the subsequent evolution necessarily depend on this spectral information. Therefore a KRL type instability, which the Morris model does not have, is required to describe this part of the evolution. It can be further conjectured that the KRL instability in conjunction with the Rayleigh model, leads to the universal self-preservation, with the Regime II value of $2\pi \delta_\omega/S$, beyond which predictions can be made with either the Rayleigh or the KRL model. But since the Morris model or any proposal that is based on the Rayleigh model alone, misses the KRL mechanism that leads to this universal  $2\pi \delta_\omega/S$; instead it leads to solutions in which both $d\delta_\omega/dt$ and  $2\pi \delta_\omega/S$ are dependent on initial conditions. Based on the results presented in Fig.\ref{Fig8}, the important conclusion can be drawn: that \emph{the spacing between coherent structures in a turbulent free shear layer is determined by what may be called a ‘resonance’ between the Rayleigh and KRL instabilities}. 

\subsection{Comments on the monopole instability and a simple merger model}
\begin{figure}
	\centerline{\includegraphics[width=6.5in]{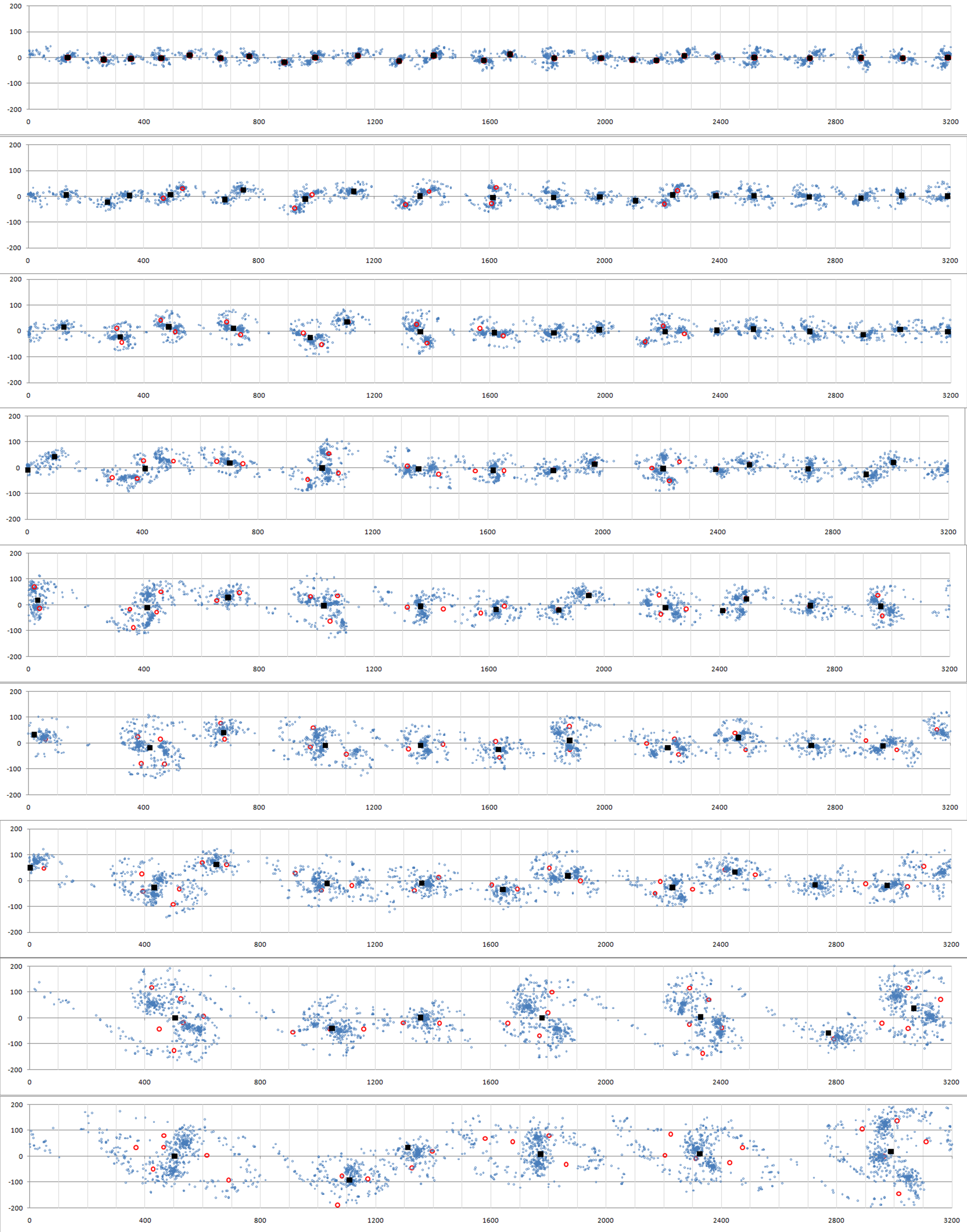}}
	\caption{\label{Fig9} Evolution of the full vortex gas solution in Regime II (blue dots) compared with the evolution of monopoles with (black squares) and without (red circles) the merger model.}
\end{figure}

As with a modal development purely based on the Rayleigh model, a pure monopole model is also incomplete in describing the entire shear layer evolution because it fails during the mergers of coherent structures. This is evident in Fig. \ref{Fig4}. Simple merger models, however, have previously been proposed(e.g. PK Dutta\cite{dutta1988discrete}).  Here a simple model is considered in which two monopoles of strength $\Gamma_1$ and $\Gamma_2$ located at $x_1$ and $x_2$ respectively, are replaced with a single monopole of strength $\Gamma_1+\Gamma_2$ located at $(\Gamma_1 x_1+\Gamma_2 x_2)/(\Gamma_1+\Gamma_2)$, when $|x_1-x_2 |<(C(\Gamma_1+\Gamma_2 ))/2\Delta U$. The only adjustable parameter in this model is $C$. From the two structure merger results described in Sec. 4 (Fig.\ref{Fig4}), a value of $C=0.7$ is a reasonable first choice.
With this model it is of interest to consider the evolution well within the universal self-preservation regime. Individual structures are replaced by monopoles with a circulation equal to that of the sum of the respective structures (i.e. with different monopole strengths) and then the full vortex-gas system and the monopole system are independently integrated beyond this point. The results for the full vortex-gas simulation, namely the basic monopole system without mergers and the monopole system with the simple merger model, are shown in Fig. \ref{Fig9}.
It can be seen that the monopoles without the merger model provide short time predictions of the evolution of structures in the self-preservation regime. (The agreement of the KRL growth exponents in Fig. \ref{Fig6}, for long waves that correspond to scenarios in which spacing between coherent structures is larger than their size, is consistent with this observation.)After the first set of mergers, however, the monopoles no longer track the evolution of the subsequent structures. (Therefore the monopole model by itself is not self-preserving. Note however, that after a sufficiently long time, the monopoles will begin to cluster once again, and Regime II would be eventually recovered, as shown by SNH). 
The introduction of the simple merger model ensures the correct description of the entire evolution, even after 3 sets of mergers. This model works in both regime Ib and II, for long times. Unlike the Rayleigh or KRL model, this monopole merger model is nonlinear and involves an adjustable constant, but it appears to be the simplest reduced-order model that describes the entire, post-coherent-structure evolution of the shear layer. It also highlights the basic instability processes that determine shear layer growth. It captures the essence of the `reverse cascade' to larger and larger scales which provides both chaotic behavior and a universal growth rate independent of initial conditions. 

\section{Conclusions}
\label{sec6}
The results from single mode analyses compare and contrast the Rayleigh and Kelvin-Biot-Savart approaches to instability. They demonstrate the equivalence of the two approaches for the linear evolution (Regime Ia), despite the different nature of the approximations involved. The agreement with the Rayleigh solution, while not unexpected, is nevertheless remarkable because the same Biot-Savart formulation is subsequently used to study the formation of coherent structures, their interaction, and eventually a universal self-preservation. Furthermore, the agreement between the vortex-gas and linearized Euler equations during the linear evolution emphasizes the important fact that, for inviscid flow, all of the dynamics is contained in the simple Kelvin, Biot-Savart mechanics.

The nonlinear saturation of the initially growing modes leads to the formation of coherent structures. It is found that the immediate evolution following the emergence of coherent structures (Regime Ib) can also be described by a linear theory - not the Rayleigh model for the growth of modes in a uniform-in-$x$ base flow, but the Karman, Rubach \& Lamb (KRL) model for the linear instability of monopoles. Once the initial growing mode is saturated and the vorticity is located in coherent structures of equal (or nearly equal) strength,the instability of the resulting array is well described by a distribution of equal-strength monopoles as in the KRL instability.  The basis of this instability is that any perturbation to either the strength or location of the monopoles in the array is unstable because the velocity of at least one vortex due to all of the vortices to the left will not be exactly equal and opposite to the velocity induced by all the vortices to the right of it.  Such a vortex will move in a way the resulting motion increases the imbalance.	

Subsequently the layer evolves to a self-preserving reverse cascade (Regime II) of SNH characterized by a universal growth rate and spectral content. Surprisingly, both Rayleigh and KRL models (for large wavelengths)are found to provide good local predictions for the growth of perturbations. In this universal regime, if the mean vorticity profile and the linear instability of this profile are assumed, and there is an energy coupling between the modes and the mean profile as in the model of Morris \emph{et al.}, it is found that it is possible to obtain the universal self-preservation, observed in the full simulations, only if a specific choice of the initial conditions is made in the Morris model. The model therefore leads to a self-preservation growth rate and spectrum that is initial-condition dependent and it therefore fails in general to predict the universality observed in the full simulations.  One possible reason for this failure is that predictions of the Rayleigh theory are valid only at a special value for the ratio of the average coherent structure spacing to layer thickness attained in Regime II.  Interestingly, this value is found to be close to the intersection of the maximum growth rates given by the linear theories of Rayleigh (assuming the mean vorticity distribution) and KRL (assuming arrays of coherent structures of uniform strength and equal spacing). This suggests the possibility of some locking between the two. It is also shown that, if in this universal regime the coherent structures are replaced by monopoles, reasonable short term predictions are found for different values of mean thickness to coherent structure spacing ratios. However, this simplification fails during merger between coherent structures. The addition of a simple semi-empirical model for mergers leads to good predictions of the long-term behavior observed in the full simulations in all regimes following the formation of the first coherent structures. This monopole merger model captures the essence of the basic instability, which leads to a `reverse cascade' to larger and larger scales and to order and disorder implied in the universal growth rate and self-preservation with a statistical distribution of monopole strengths and spacings.	The simplicity of the Kelvin Biot-Savart formulation in this two-dimensional inviscid case and the connection that has been shown between the linear instability theory for the mean vorticity profile and the monopole instability of coherent structures, illustrate both the chaotic and structured behavior of the vorticity field in this simple case.  Such behavior may be broadly characteristic of many more complex turbulent shear flows. Preliminary analyses suggest that the Kelvin Biot-Savart formulation is perhaps of more general applicability to instability in shear flows, and can be used to describe evolution of 2D Bickley jets\cite{SuryanarayananBrown2011} and viscosity stratified flows (Suryanarayanan, Tendero Ventanas, Govindarajan and Theofilis, \textit{manuscript under preparation}).

\section{Acknowledgments}
We dedicate this paper to the memory of Prof.Anatol Roshko, with whom we have had the pleasure of having many rewarding discussions. We thank Prof. Rama Govindarajan (ICTS Bangalore) and Dr.Sourabh Diwan (IISc)for discussions. We gratefully acknowledge Prasanth P (JNCASR) for providing DNS calculations for comparison.  We acknowledge support from DRDO through Project No. RN/DRDO/4124 and from Intel through Project No. RN/INTEL/4288. Garry Brown wishes to acknowledge the support from the Jawaharlal Nehru Centre for Advanced Scientific Research and travel support from the Indo-US Science and Technology Forum (IUSSTF).  

\appendix

\section{The point-vortex `eigen mode'}

The Rayleigh solution for the instability of a piecewise linear velocity profile (a uniform vorticity shear layer of finite thickness $2\delta$) is a pair of vortex sheets at $y=\pm \delta$, whose strengths grow exponentially with time. Due to the linear approximation, the vorticity does not actually 'move' in the Rayleigh solution. In the point-vortex solution however, the vorticity crosses  $y=\pm \delta$ and does so even at very early times.  That is, in the point-vortex system, during the initial stage of exponential growth of perturbations, there is a sinusoidal perturbation of the outer row of vortices and also of every initial row, but with varying amplitude and phase. The inter-vortex spacing on each initial row also varies approximately sinusoidally with $x$, with its amplitude and phase varying for each row.  While this may appear at first sight as if a vorticity perturbation is generated within the strip, the ‘coarse-grained vorticity’, averaged over the area of the initial cell, is still conserved. The reason is that the combination of an increase in the local concentration of vortices along a row is matched by a local increase in distance between rows and this leads to a negligible change in vortices per unit area. This is true because the velocity field is the curl of a vector potential, from which the Biot-Savart relationship is found, and hence the velocity field is divergence-free. The shape of the area gets distorted with time. 	

\begin{figure}
	\centerline{\includegraphics[width=5.0in]{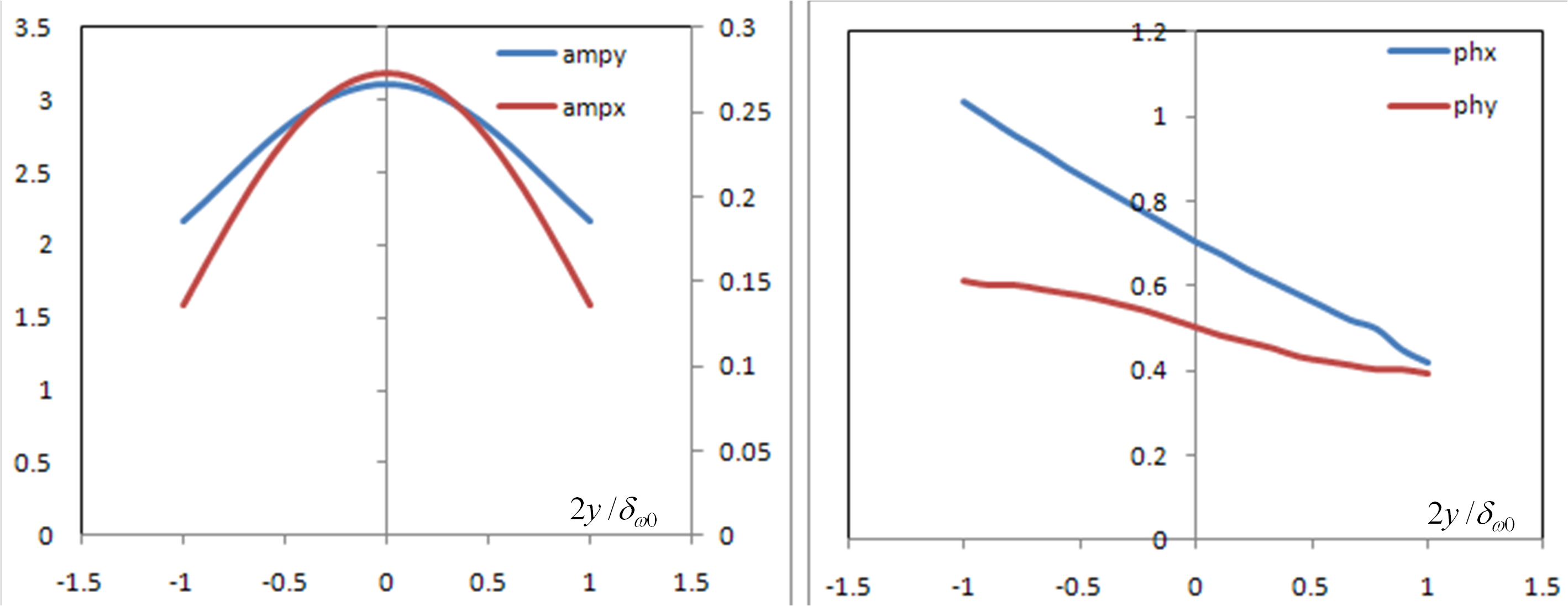}}
	\caption{\label{FigA1} The ‘numerically obtained eigen mode for vortex spacing’.ampx, ampy denote the perturbation amplitudes of the $x-$spacing and $y-$displacements from the base configuration, and phx and phy denote the respective phases.  }
\end{figure}

\begin{figure}
	\centerline{\includegraphics[width=5.0in]{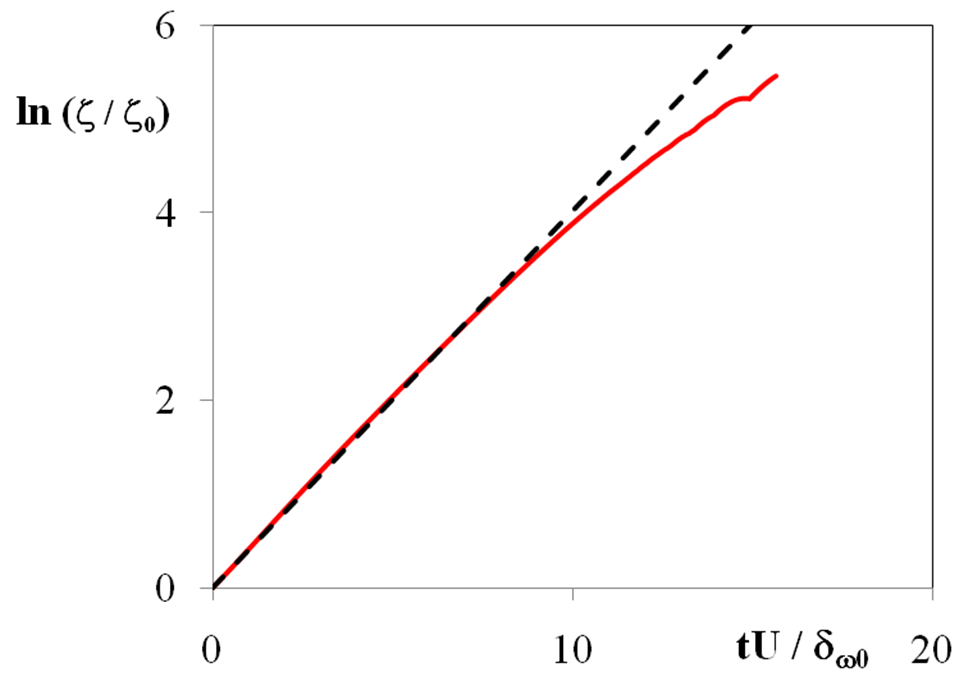}}
	\caption{\label{FigA2} Exponential growth from $t = 0$ with the `eigen mode' initial condition. The growth rate given by Rayleigh (0.402) is shown in dashed lines.}
\end{figure}

As a consequence, the vorticity eigenmode of the point-vortex system has a subtle difference from that of linear stability theory, though it could be expected to be very close in the small amplitude limit.  In order to understand this better, we repeated the simulation with the same number of vortices, but initialized the displacements of the vortices from their regular undisturbed positions with a `numerically obtained eigen mode'.  This was obtained by (i) `rescaling' the observed exponentially growing $x$ and $y$ perturbations in the position of vortices in each row found in the linear (exponential) regime of the simulation shown in Fig.\ref{Fig2} and (ii) then reducing the perturbations exponentially to a much earlier time when the normal displacement amplitude on the central row was only $0.0005$ of the wavelength (c.f. initially $0.05$).  With this initial condition, exponential growth was found from very nearly $t=0$, as seen in Fig.\ref{FigA2}, and the least-square fitting of the growth rate agrees with the exact solution to three significant digits. This procedure, therefore, gives an approximation to the actual eigen function for the distribution of the vorticity (Fig. \ref{FigA1}); this result is obtained here, in the absence of a theory, through the successive application of the Biot-Savart relationship.

\section{Convergence, Chaos and Desingularization}

It is well known that more than three point vortices (in an infinite plane) can become chaotic \cite{aref1983integrable,novikov1978stochastic}. The emergence of irregular behavior in the roll up of a vortex sheet, represented by a single row of point vortices, has been studied in detail\cite{birkhoff1962helmholtz,moore1979spontaneous,krasny1986study}.  The problem is less severe in the evolution of a finite thickness shear layer represented by multiple rows of point vortices, as the correct large scale behavior is recovered. However it can be observed that the motion of some point vortices became chaotic in local regions during the non-linear evolution of the present system. 

\begin{figure}
	\centerline{\includegraphics[width=5.0in]{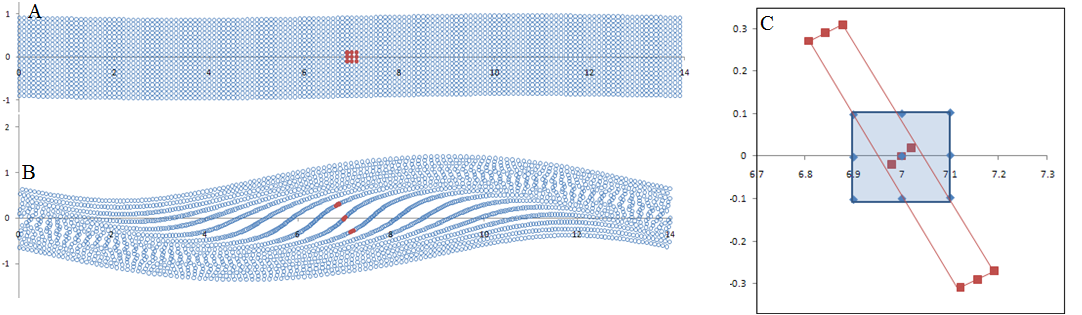}}
	\caption{\label{FigB1} `Ill-conditioning' due to the straining motion on the initial elemental area whose uniform vorticity distribution is represented by a delta function.}
\end{figure}

Figure \ref{FigB1} explains a possible origin for this behavior. The original area in the initial distribution of point vortices shown in Fig.\ref{FigB1}(a) , and is represented by the blue square in  Fig. \ref{FigB1}(c). The area becomes distorted by the straining motion into an approximate rectangle with a high aspect ratio, as in Figures \ref{FigB1}(b) and \ref{FigB1}(c).  The single point vortex at the center becomes much closer to that representing the vorticity in the adjacent area.  Thus some of the vortices, that were initially in the same row, move toward each other and away from adjacent rows. It is easy to understand that the closer they are the more susceptible they would become to a local rotation about each other, which could be expected to lead to locally chaotic behavior.  The method conserves the area of an element of fluid (by continuity), but the shape of the area can become greatly distorted.  It can be seen that the place where chaos emerges is the place where the elements at that time become most `ill-conditioned'.

Another way of thinking about the phenomenon is that when the distance between rows is large and the distance between neighboring vortices in a row is small, as it is when the area becomes ill-conditioned, the row increasingly behaves locally like an isolated vortex sheet.  It is then subject to the instability of a sheet, namely it becomes susceptible to the smallest wavelength perturbation. Hence it may suffer instability at the scale of inter-vortex spacing, leading to a local roll up and subsequent chaotic behavior. Hence the Krasny-Majda desingularization method, which was developed for a sheet, might also be expected to delay chaos in the present system.

\begin{figure}
	\centerline{\includegraphics[width=5.0in]{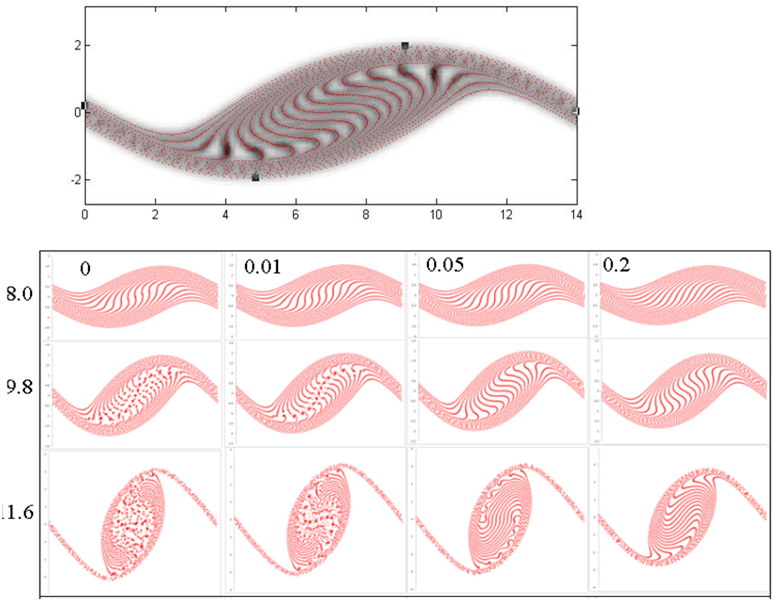}}
	\caption{\label{FigB2} The effect at different times of increasing the Krasny desingularization parameter.   The values of the desingularization `core radius', with respect to the initial inter-vortex spacing, are 0,1.4, 7 and 28 and 0, 0.078, 0.389 and 1.55 in terms of $\delta_{\omega 0}$}
\end{figure}

Figure \ref{FigB2} shows the evolution of instability with  identical initial conditions for different desingularization parameters. As illustrated, desingularization delays the onset of chaos by limiting the local velocity between neighboring vortices.  This leads evidently to a ‘better’ local representation of the continuous (constant) vorticity field.  The local ‘sheets’ are now of finite thickness and hence are not unstable to small wavelength perturbations. But this only delays the onset of chaos, as eventually disturbances of wavelengths that are long enough to destabilize the local finite-thickness sheet, emerge. Further, when the desingularization radius becomes larger than approximately 15\% of the fundamental wavelength, the large scale behavior in these calculations becomes noticeably affected. 

\begin{figure}
	\centerline{\includegraphics[width=5.0in]{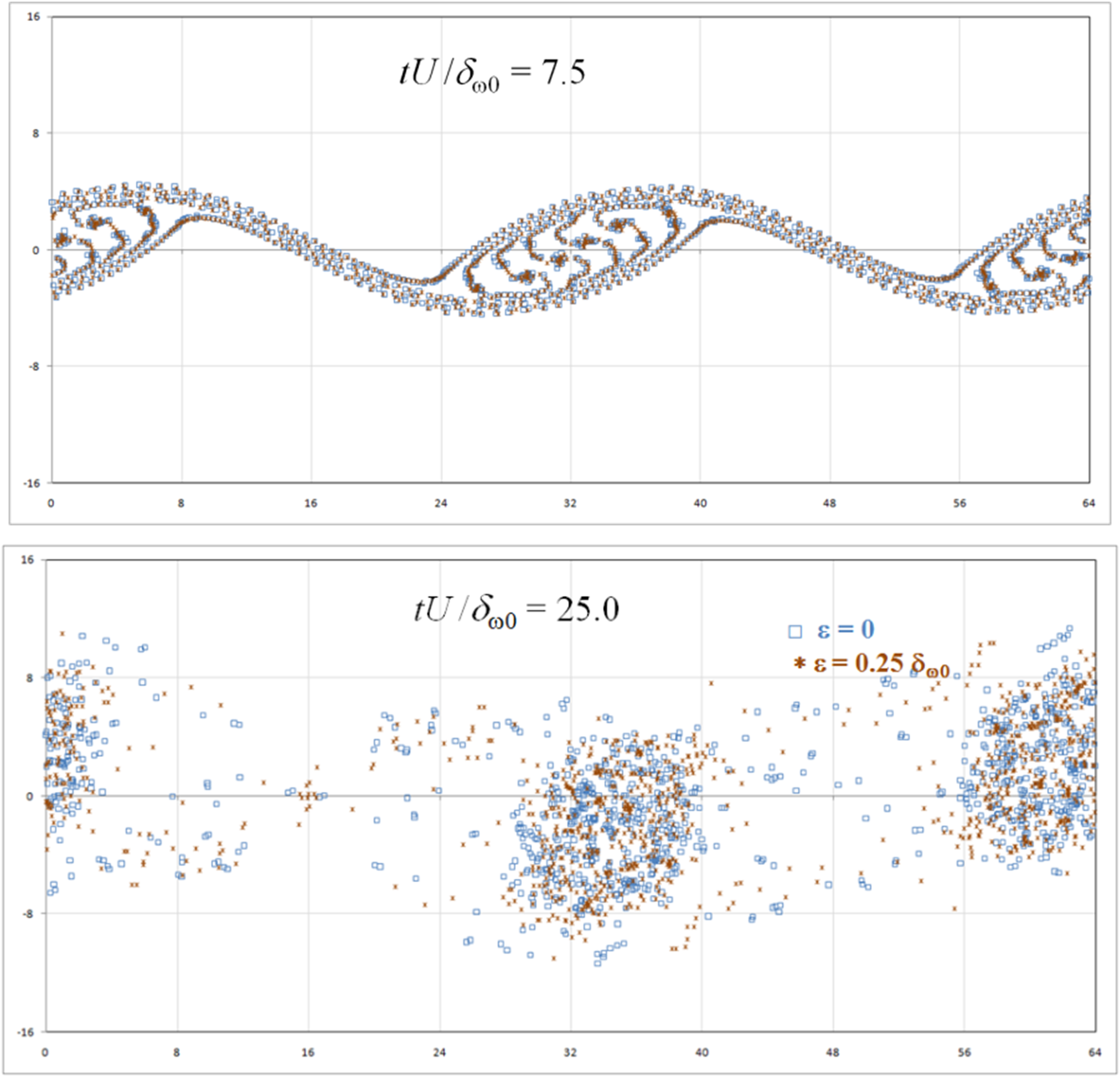}}
	\caption{\label{FigB3} Evolution of a shear layer initialized with a fundamental and subharmonic with and without desingularization.}
\end{figure}

As seen in Figure \ref{FigB2}, the large scale development associated with the roll-up of the fundamental appears essentially unaffected by desingularization up to a core radius of at least 7 with respect to the initial vortex spacing (5\% of the instability wavelength).  However, the point vortex chaos can act as a weak `random forcing’ to the `monopole' locations, and this introduces a weak sub-harmonic forcing that can trigger the subsequent evolution (i.e. Regime I(b)), even when no sub-harmonic perturbation was present in \emph{the initial condition}. However, as long as a subharmonic is explicitly introduced in the initial condition, the subsequent evolution (in a coarse-grained sense) is unaffected by desingularization as seen in Fig. \ref{FigB3}.

A related question is on the role of chaos and the convergence of the solution with increasing $N$, ultimately to a solution of the 2D Euler equations (i.e to the evolution of the continuous vorticity field).  There have been many theorems on convergence of vortex methods. Hald\cite{hald1979convergence} proposed that the vortex blob method converges to the smooth solution of the Euler equations in the limit $N \rightarrow \infty$ but only when the inter vortex spacing goes to zero faster than the blob radius, implying that the blobs must overlap. This means that the true point vortex methods (i.e $\epsilon= 0$) do not converge to the continuous solution. It was later shown that the convergence criterion is influenced by the choice of the norm, and Goodman\cite{goodman1990convergence} showed that the blob method converges in the limit of point vortices provided the correct norm is adopted. Further discussion  including proofs on the point vortex method providing a the weak solution of the Euler equations are presented by Marchioro \& Pulvirenti\cite{marchioro1993}.

\begin{figure}
	\centerline{\includegraphics[width=3.5in]{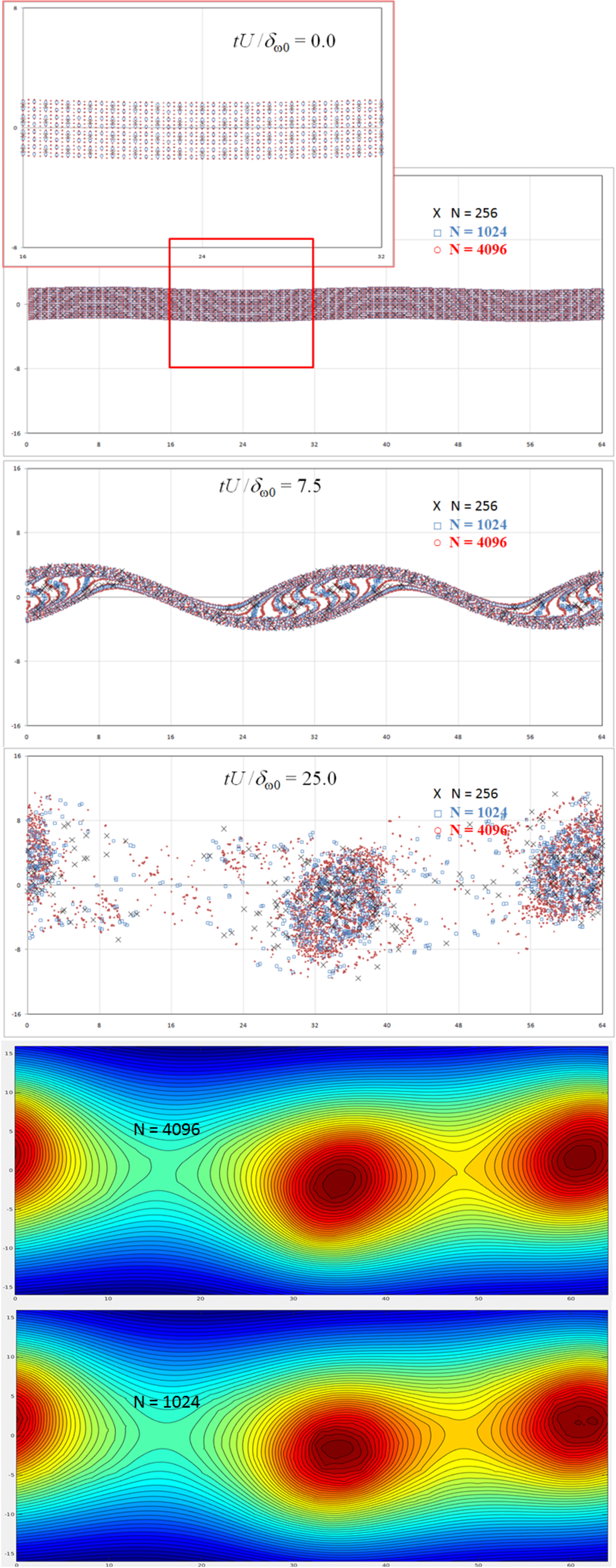}}
	\caption{\label{FigB5} (a) - (c) time evolution of a finite thickness shear layer initialized with fundamental and sub-harmomic discretized by 256, 1024 and 4096 vortices.  (d) and (e) - contour plot of streamfunction at $tU/\delta_{\omega 0}= 25$ for $N = 1024$ and $4096$.  }
\end{figure}

\begin{figure}
	\centerline{\includegraphics[width=6.5in]{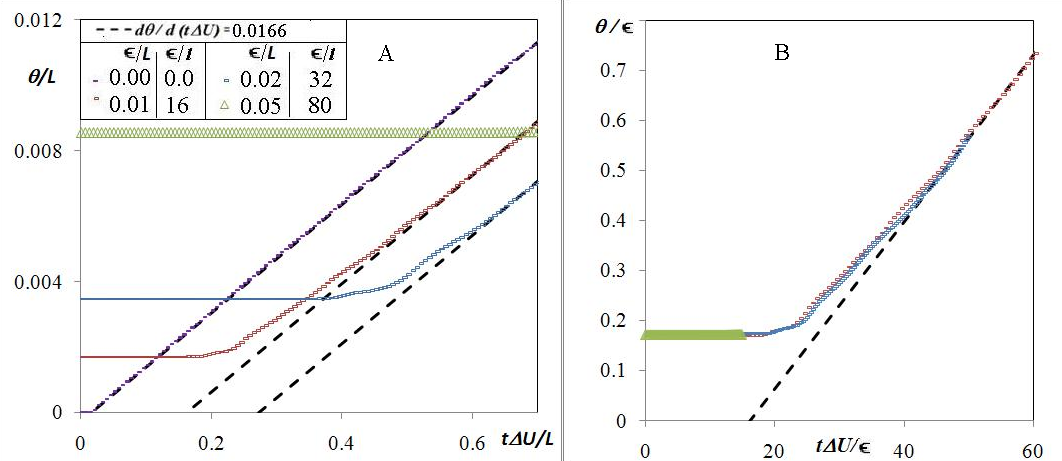}}
	\caption{\label{FigB4} \textbf{A.} Increase in $\epsilon$ delays (in terms of $t\Delta U/L$ and $t\Delta U/l$) the onset of Regime II, but has no influence on the spread rate in Regime II - The observation on universality in the non-equilibrium evolution is unaffected on desingularizing the vortices (with any given $\epsilon/l$ at sufficiently large $N$ ). \textbf{B.} Scaling based on $\epsilon$, (note that $\theta/\epsilon = F [t\Delta U/\epsilon]$ for $\epsilon>>a$ and $t\Delta U /L << 1$) showing agreement between the different datasets. }
\end{figure}

Without getting into an elaborate discussion of the mathematical theorems, we present a numerical demonstration of convergence. In the finite-thickness shear-layer evolution considered earlier, the solution converges in a coarse grained sense with increase in $N$, as seen in Fig. \ref{FigB5}. Individual vortex motion is chaotic beyond a certain critical time that may depend on $N$, and hence there will not be a one to one correspondence at the level of individual vortices, but the coarse-grained / large-scale solution can be observed to converge with $N$. This is clearer from the stream function which is obtained from the vector potential.As it is derived from the vorticity field, it is still technically chaotic, but being an integral quantity, it is dominated by the long range influence of the large scale vorticity distribution rather than by locally chaotic motion. We are interested in the large scale motion at the level of coherent structures, because once the coherent structures form their subsequent interaction is initially described by the instability of monopoles (see Sec.\ref{sec4}), i.e. it does not depend on the detailed vorticity distribution within the coherent structure. The solution is eventually chaotic at the level of coherent structures, but once again the detailed structure does not matter owing to the statistical nature of the evolution (elaborated in the following paragraph). Since the solution is seen to converge with $N$ and is independent of desingularization at the level of large scale description (Fig.\ref{FigB4}), the present results, taken together with Appendix C, can be considered as having general applicability and relevant to the solution of the 2D Euler equations. 

Thus the results on convergence, chaos and desingularization in the context of the results presented in this paper can be summarized as follows.The origin of chaos during the initial roll up into structures (i.e. Regime RIa) may be dependent on factors such as $N$ and $\epsilon$.  But once the structures roll up, all cases (including the continuous vorticity case)essentially transform into a system of monopoles, where each coherent structure behaves as a vortex. Such a system is necessarily chaotic (regardless of the numerical method) as long as there is a sufficiently large number of coherent structures in all. So once there is chaos at the level of the coherent structures, even an infinitesimal noise (which may arise from free stream turbulence, wind tunnel noise, thermal noise due to molecular motion,  numerical round-off errors that exist in all numerical methods) will lead to an initial-condition and stochastic forcing independent statistical solution.  In mechanistic terms this can be described in the following way. Once there are coherent structures (which are achieved by roll up of the initial instability waves), even if they are all ordered, their subsequent motion is very sensitive to the smallest perturbations in the system. As a result, the smallest perturbation will cause them to migrate from their unstable equilibria and the subsequent chaos will lead them to merge unevenly leading to a next generation of coherent structures that have different strengths and different locations, even if starting with equi-spaced, equal strength structures.  The process continues till statistics become independent of the initial conditions or the noise in the system. So while the exact source of initial noise or stochastic forcing may be different for each method (e.g. continuous vorticity on discrete grid vs. discrete vortex method, wind tunnel vs. numerical simulations),  and such a forcing may be essential,  the answer beyond a point does not depend on the precise nature of the forcing.  

The above statements are be supported by the following different kinds of evidences. 
(1) The same universality is obtained using different initial conditions and different $N$ \cite{snh}
(2) The same universality for different values of the desingularization parameter that suppresses small scale chaos (Fig. \ref{FigB4}).
(3) The same universality is obtained when a deliberate stochastic `forcing' is introduced via adding a random walk component to vortex motion\cite{suryanarayanan2014insights}.
(4) Most high Reynolds number experiments and DNS are in close agreement with the universal spread rate observed in RII of point-vortex shear layers \cite{suryanarayanan2017insights}.

\section{Results for growth exponents for an initially `thick' shear layer}
In this Appendix the results of a simulation with an initially `thick' shear layer are presented. To generate the initial condition for this simulation 4480 vortices arranged in 4 rows (i.e. $L/l = 1120$), such that they are equi-spaced in both $x$ and $y$. A uniformly distributed random number (with $a = 0.5l$) is then added to $y-$ positions of each column of vortices. We use a Krasny-type desingularization for these simulations, with $\epsilon/l$ = 1.  The combination of the present initial condition and desingularization results in an initial vorticity thickness $\delta_{\omega0}=4l$ . The results are then averaged over 16 realizations.  

\begin{figure}
	\centerline{\includegraphics[width=5.0in]{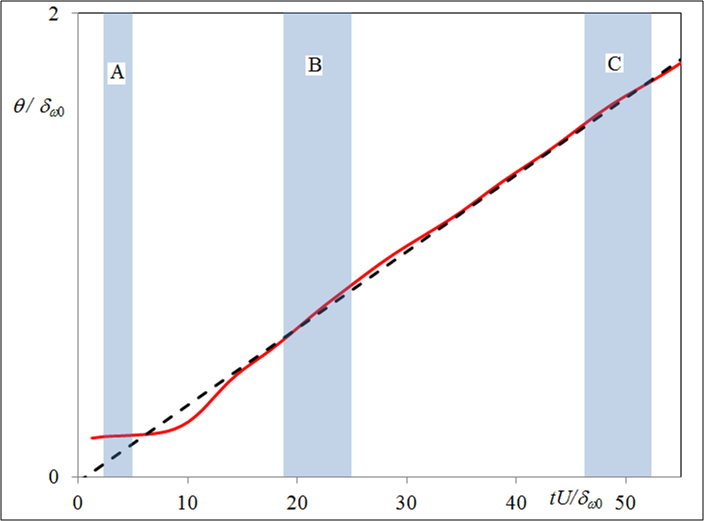}}
	\caption{\label{FigC1} Evolution of momentum thickness of vortex gas (thick) shear layer.  The dashed line shows the universal spreading rate ($d\theta / d(t\Delta U) = 0.0166$ (SNH)) in Regime II.  Region A corresponds to times when the base flow (thickness) hardly changes, region B indicates a region where the layer grows rapidly but not at the universal rate, and region C when the layer grows at the universal rate. The respective spectral evolutions are analyzed and presented in Figures \ref{FigC2}, \ref{FigC3} and \ref{FigC4} respectively.  }
\end{figure}

\begin{figure}
	\centerline{\includegraphics[width=6.5in]{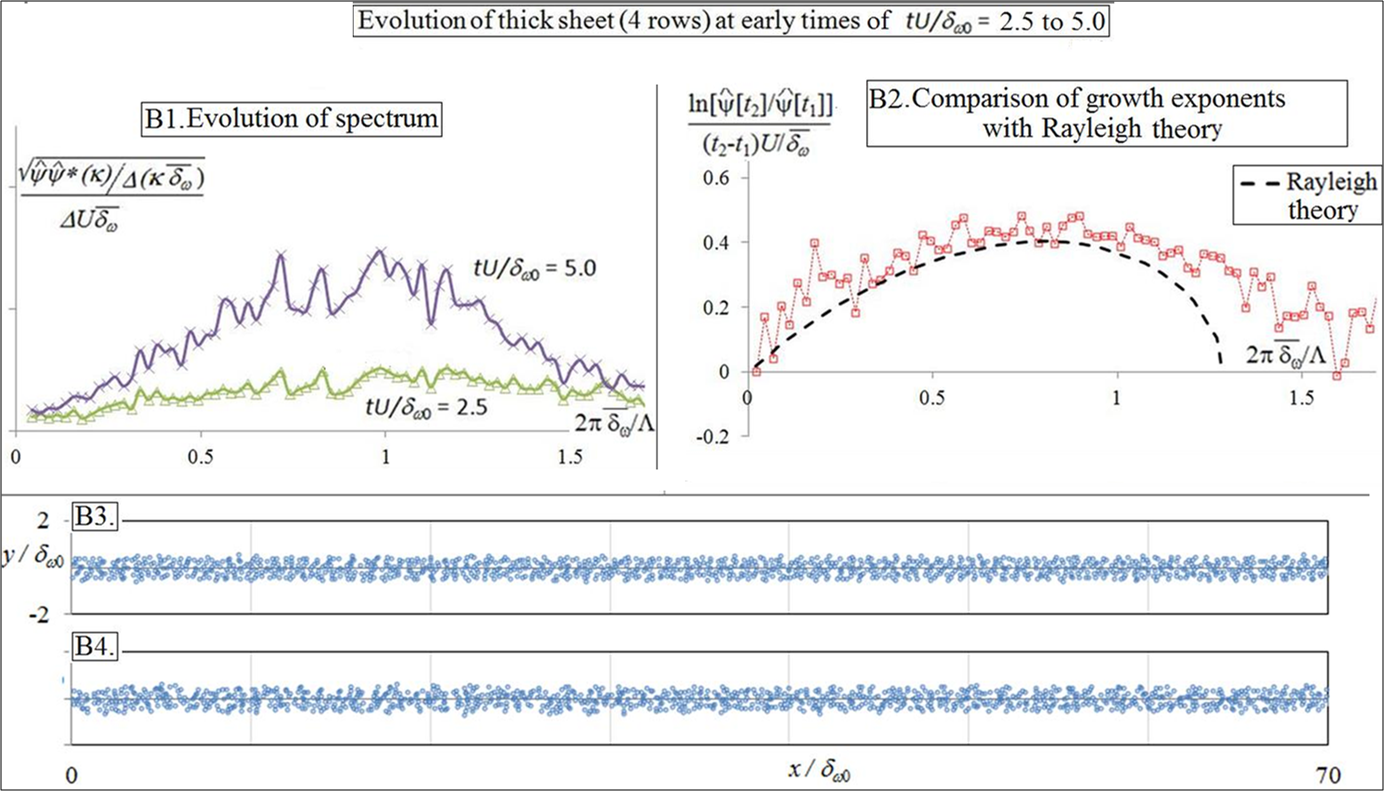}}
	\caption{\label{FigC2} Spectral evolution during early times (marked as \textbf{A} in Fig.\ref{FigC1} $tU/\delta_{\omega 0} = 2.5$ to 5.0). \textbf{B1} shows the power spectral density of the perturbation stream function and \textbf{B2} shows the respective growth exponents, \textbf{B3} and \textbf{B4} show the vortex-positions at  $tU/\delta_{\omega 0} = 2.5$ and 5.0 that clearly indicate that distinct coherent structures are yet to form.}
\end{figure}

The evolution of the momentum thickness is shown in Fig.\ref{FigC1}. At small times, the thickness of the layer hardly changes. This is consistent with Regime I(a), the linear growth of initial perturbations.  The evolution of the layer from  $tU/\delta_{\omega 0} = 2.5$ to $5.0$ (marked as region A in Fig.\ref{FigC1}) is shown in Fig.\ref{FigC2}. It can be seen that the growth exponents (Fig.\ref{FigC2}.B2) agree with the predictions of Rayleigh theory, consistent with findings in Sec.\ref{sec3}. 

\begin{figure}
	\centerline{\includegraphics[width=6.5in]{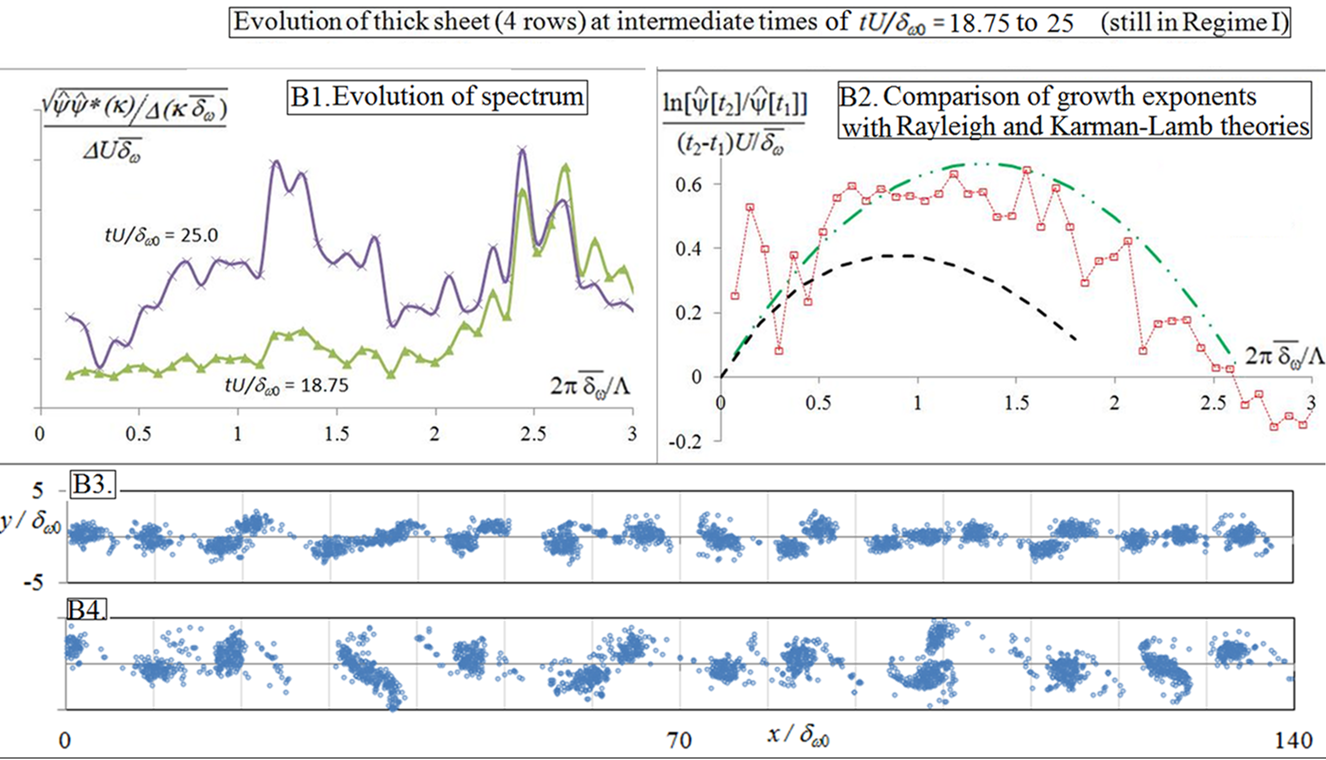}}
	\caption{\label{FigC3} Evolution during $tU/\delta_{\omega 0}$= 18.75 to 25.0 (RIb, after formation of coherent structures). $\overline{\delta_\omega}$ is the average thickness during the evolution considered. \textbf{B1} shows the power spectral density of the perturbation stream function and \textbf{B2} shows the growth exponents, \textbf{B3} and \textbf{B4} show the vortex-positions at  $tU/\delta_{\omega 0}$= 18.75 and 25.0. }
\end{figure}

We next consider the evolution of the layer from $tU/\delta_{\omega 0} = 18.75$ to $25$.  The thickness of the layer has rapidly increased, and is over double the initial thickness by $tU/\delta_{\omega 0} = 18.75$. The evolution of the layer during this time is shown in Fig.\ref{FigC3}. It can be observed (Fig.\ref{FigC3} B3 and B4) that there are distinct coherent structures during this phase of the evolution of the layer. The spectrum at $tU/\delta_{\omega 0} = 18.75$ has a distinct peak, and that at $tU/\delta_{\omega 0} = 25.00$ has a second peak, approximately at the `subharmonic'. The growth exponents (Fig.\ref{FigC3}B2) do not agree with the predictions of Rayleigh theory for the `base flow' but closely agree with the KRL predictions. These observations are consistent with the Regime I(b) results presented in Sec.\ref{sec4}.

\begin{figure}
	\centerline{\includegraphics[width=6.5in]{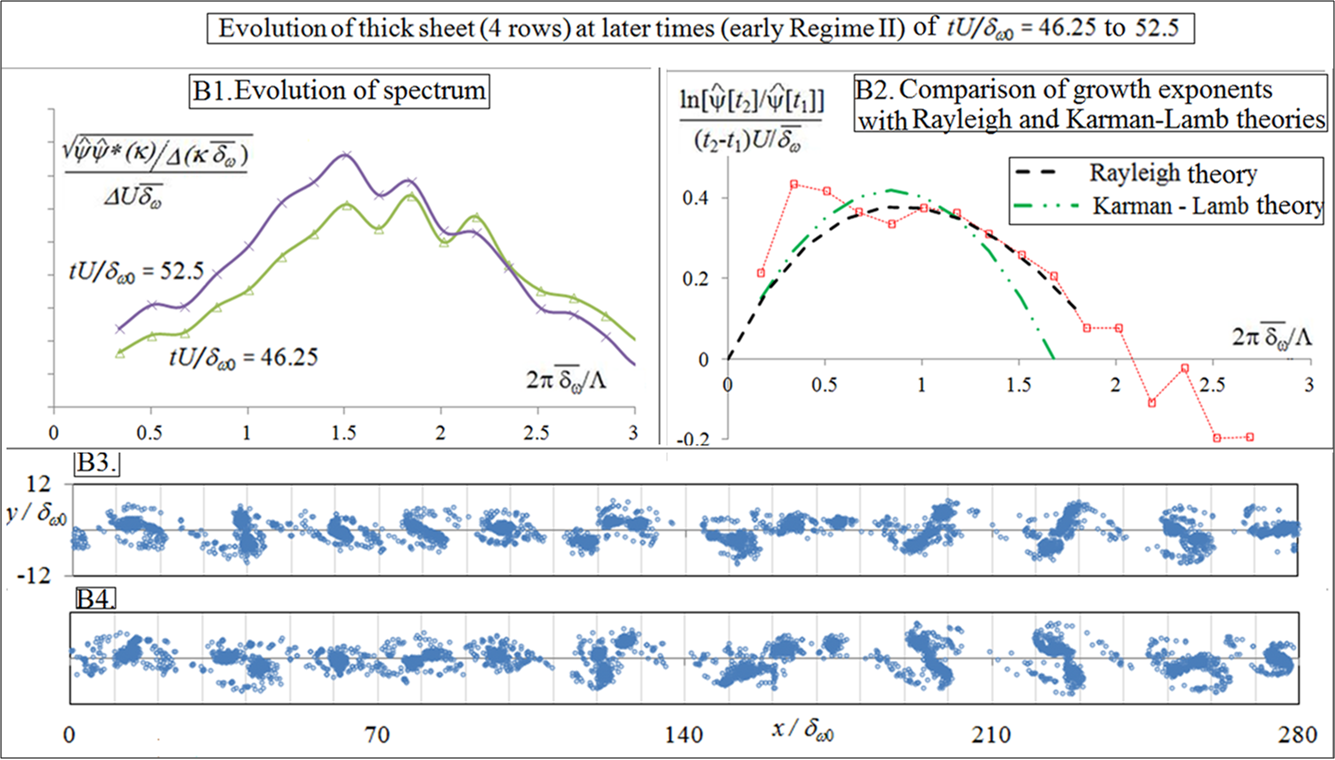}}
	\caption{\label{FigC4} Evolution during $tU/\delta_{\omega 0}$= 46.25 to 52.5 (transition to Regime II).\textbf{ B1} show the power spectral density of the perturbation stream function scaled with the time-averaged thickness, \textbf{B2} shows the growth exponents, \textbf{B3} and \textbf{B4} show the vortex-positions at  $tU/\delta_{\omega 0}$ = 46.5 and 52.5. }
\end{figure}

At later times (say from $tU/\delta_{\omega 0} = 46.25$ to $52.5$ - marked as `C' in Fig.\ref{FigC1}), the momentum thickness grows with the same universal rate as the `thin shear layer' case discussed in Sec.\ref{sec5}.  The evolution of the layer is shown in Fig.\ref{FigC4}. The results shown here, namely that the observed growth exponents agree with both Rayleigh and KRL, are consistent with the RII results discussed in Sec.\ref{sec5}. Thus all three regimes - RI(a), RI(b) and RII are recovered, and can be distinctly observed for this class of initial conditions.

\section{The Morris-model for the 2D Incompressible temporal free shear layer}

Consider perturbation modes of the form $\mathbf{\hat{u}}(\kappa,y,t)\exp(i\kappa x)$.  The perturbation velocity field in the limit of a large number of modes is

\begin{equation}
\mathbf{u'}=\delta_\omega \int d\kappa \exp(i\kappa x) \mathbf{\hat{u}}(\kappa,y,t)
\end{equation}

Note that both $u'$ and $\hat{u}$ have dimensions of velocity.

The kinetic energy per unit length in $x$ associated with the perturbations is 

\begin{equation}
E_p(t) = \frac{1}{2} \int_{-\infty}^{\infty} dy <\mathbf{u'}.\mathbf{u'}> = \frac{\delta_\omega}{2}\int d\kappa  \int_{-\infty}^{\infty} dy <\mathbf{\hat{u}}.\mathbf{\hat{u}}>
\label{eqnD0}
\end{equation}

If we choose $A$ to represent an effective amplitude, such that 

$A^2(\kappa,t)=(1/4) \int_{-\infty}^{\infty} dy  <\mathbf{\hat{u}}.\mathbf{\hat{u}}>$, eqn.\ref{eqnD0} can be written as

\begin{equation}
E_p(t) = 2 \int d\kappa \delta_\omega A^2(\kappa,t)
\label{eqnD1}
\end{equation}

Note that $A$ not only depends on the physical amplitude of the perturbation but also the distribution of the mode in $y$. The constant in front has no unique meaning, it is chosen to agree with the choice of normalization adopted by Morris \textit{et al}\cite{morris1990turbulent} .  (It can be shown that if the model were to lead to a self-similar solution independent of the initial conditions, the exact normalization is irrelevant, as it can be absorbed in the equations in such a way that it will only change the initial value of $A$).

The rate of change of the perturbation energy is given by

\begin{equation}
\frac{d E_p}{dt} = 2 \int d\kappa \big(\frac{d(\delta_\omega A^2)}{dt}\big)
\label{eqnD1p5}
\end{equation}

The Morris model assumes that the perturbation grow via linear mechanisms, i.e. $A\sim\exp(Gt)$, where $G$ is a growth exponent given by Rayleigh theory for the base flow at that instant of time.  If the base flow is frozen, 

\begin{equation}
\frac{d A^2}{dt} = 2G A^2
\end{equation}

But the Morris model also accounts for the growth of the base flow. Therefore

\begin{equation}
\frac{d}{dt} (A^2 \delta_\omega) = 2G (A^2\delta_\omega)
\label{eqnD2}
\end{equation}

Thus eq.\ref{eqnD1p5} can be written as

\begin{equation}
\frac{d E_p}{dt} = 4\delta_\omega \int d\kappa A^2(\kappa,t) G(\kappa)
\label{eqnD3}
\end{equation}

The change in the kinetic energy of the base flow is
\begin{equation}
\frac{d}{dt} (E_m(t)) = \frac{d}{dt} \big(\frac{1}{2} \int_{-\infty}^{\infty} dy {\overline{U}}^2 \big) = \int_{-\infty}^{\infty} dy  \overline{U} \frac{d\overline{U}}{dt}
\end{equation}

where $ \overline{U} = \Delta U F(\eta \equiv \frac{y}{\delta_\omega(t)})$

\begin{equation}
\frac{d}{dt} (E_m(t)) = -\frac{d\delta_\omega}{dt} \Delta U^2 \int_{-\infty}^{\infty} d\eta \eta FF'
\end{equation}

For a tanh profile with unit velocity difference and unit vorticity thickness, the integral on the right hand side is equal to $1/8$. Thus 

\begin{equation}
\frac{d}{dt} (E_m(t)) = -\frac{\Delta U^2}{8} \frac{d \delta_\omega}{dt}
\label{eqnD4}
\end{equation}

Since the total kinetic energy is conserved for the incompressible flow, 

\begin{equation}
\frac{d}{dt} (E_m(t)+E_p(t)) = 0
\label{eqnD5}
\end{equation}

Substituting eq.\ref{eqnD3} and eq.\ref{eqnD4} in eq.\ref{eqnD5}, 

\begin{equation}
4\delta_\omega \int d\kappa A^2(\kappa,t) G(\kappa) -\frac{\Delta U^2}{8} \frac{d \delta_\omega}{dt} = 0
\end{equation}

\begin{equation}
4\delta_\omega \int d\kappa A^2(\kappa,t) G(\kappa) -\frac{\Delta U^2}{8} \frac{d \delta_\omega}{dt} = 0
\end{equation}

Rearranging, we get the growth rate of the layer as 

\begin{equation}
 \frac{d \delta_\omega}{dt} = \frac{32}{\Delta U^2}\delta_\omega \int d\kappa A^2(\kappa,t) G(\kappa)
 \label{eqnD7}
\end{equation}

Rewriting eqn.\ref{eqnD2}, we obtain the following equation for the growth of the normalized amplitude of each mode as 

\begin{equation}
\frac{d}{dt} (A^2 (\kappa,t) ) = A^2(\kappa,t) \big(2G(\kappa)  - \frac{1}{\delta_\omega} \frac{d\delta_\omega}{dt}\big)
\label{eqnD8}
\end{equation}

where the growth exponents $G(\kappa)$ are obtained from linear stability theory (Rayleigh). Equations \ref{eqnD7} and \ref{eqnD8} can be solved as an initial value problem in time.

\bibliography{Stability-references}

\begin{thebibliography}{44}%
\makeatletter
\providecommand \@ifxundefined [1]{%
 \@ifx{#1\undefined}
}%
\providecommand \@ifnum [1]{%
 \ifnum #1\expandafter \@firstoftwo
 \else \expandafter \@secondoftwo
 \fi
}%
\providecommand \@ifx [1]{%
 \ifx #1\expandafter \@firstoftwo
 \else \expandafter \@secondoftwo
 \fi
}%
\providecommand \natexlab [1]{#1}%
\providecommand \enquote  [1]{``#1''}%
\providecommand \bibnamefont  [1]{#1}%
\providecommand \bibfnamefont [1]{#1}%
\providecommand \citenamefont [1]{#1}%
\providecommand \href@noop [0]{\@secondoftwo}%
\providecommand \href [0]{\begingroup \@sanitize@url \@href}%
\providecommand \@href[1]{\@@startlink{#1}\@@href}%
\providecommand \@@href[1]{\endgroup#1\@@endlink}%
\providecommand \@sanitize@url [0]{\catcode `\\12\catcode `\$12\catcode
  `\&12\catcode `\#12\catcode `\^12\catcode `\_12\catcode `\%12\relax}%
\providecommand \@@startlink[1]{}%
\providecommand \@@endlink[0]{}%
\providecommand \url  [0]{\begingroup\@sanitize@url \@url }%
\providecommand \@url [1]{\endgroup\@href {#1}{\urlprefix }}%
\providecommand \urlprefix  [0]{URL }%
\providecommand \Eprint [0]{\href }%
\providecommand \doibase [0]{http://dx.doi.org/}%
\providecommand \selectlanguage [0]{\@gobble}%
\providecommand \bibinfo  [0]{\@secondoftwo}%
\providecommand \bibfield  [0]{\@secondoftwo}%
\providecommand \translation [1]{[#1]}%
\providecommand \BibitemOpen [0]{}%
\providecommand \bibitemStop [0]{}%
\providecommand \bibitemNoStop [0]{.\EOS\space}%
\providecommand \EOS [0]{\spacefactor3000\relax}%
\providecommand \BibitemShut  [1]{\csname bibitem#1\endcsname}%
\let\auto@bib@innerbib\@empty
\bibitem [{\citenamefont {Darrigol}(2005)}]{darrigol2005worlds}%
  \BibitemOpen
  \bibfield  {author} {\bibinfo {author} {\bibfnamefont {O.}~\bibnamefont
  {Darrigol}},\ }\href@noop {} {\emph {\bibinfo {title} {Worlds of flow: A
  history of hydrodynamics from the Bernoullis to Prandtl}}}\ (\bibinfo
  {publisher} {Oxford University Press},\ \bibinfo {year} {2005})\BibitemShut
  {NoStop}%
\bibitem [{\citenamefont {Brown}\ and\ \citenamefont
  {Roshko}(1974)}]{brown1974density}%
  \BibitemOpen
  \bibfield  {author} {\bibinfo {author} {\bibfnamefont {G.~L.}\ \bibnamefont
  {Brown}}\ and\ \bibinfo {author} {\bibfnamefont {A.}~\bibnamefont {Roshko}},\
  }\bibfield  {title} {\enquote {\bibinfo {title} {On density effects and large
  structure in turbulent mixing layers},}\ }\href@noop {} {\bibfield  {journal}
  {\bibinfo  {journal} {Journal of Fluid Mechanics}\ }\textbf {\bibinfo
  {volume} {64}},\ \bibinfo {pages} {775--816} (\bibinfo {year}
  {1974})}\BibitemShut {NoStop}%
\bibitem [{\citenamefont {Malkus}(1956)}]{malkus1956outline}%
  \BibitemOpen
  \bibfield  {author} {\bibinfo {author} {\bibfnamefont {W.}~\bibnamefont
  {Malkus}},\ }\bibfield  {title} {\enquote {\bibinfo {title} {Outline of a
  theory of turbulent shear flow},}\ }\href@noop {} {\bibfield  {journal}
  {\bibinfo  {journal} {Journal of Fluid Mechanics}\ }\textbf {\bibinfo
  {volume} {1}},\ \bibinfo {pages} {521--539} (\bibinfo {year}
  {1956})}\BibitemShut {NoStop}%
\bibitem [{\citenamefont {Roshko}(2000)}]{roshko2000problem}%
  \BibitemOpen
  \bibfield  {author} {\bibinfo {author} {\bibfnamefont {A.}~\bibnamefont
  {Roshko}},\ }\bibfield  {title} {\enquote {\bibinfo {title} {On the problem
  of turbulence},}\ }\href@noop {} {\bibfield  {journal} {\bibinfo  {journal}
  {Current Science}\ ,\ \bibinfo {pages} {834--839}} (\bibinfo {year}
  {2000})}\BibitemShut {NoStop}%
\bibitem [{\citenamefont {Gaster}, \citenamefont {Kit},\ and\ \citenamefont
  {Wygnanski}(1985)}]{gaster1985large}%
  \BibitemOpen
  \bibfield  {author} {\bibinfo {author} {\bibfnamefont {M.}~\bibnamefont
  {Gaster}}, \bibinfo {author} {\bibfnamefont {E.}~\bibnamefont {Kit}}, \ and\
  \bibinfo {author} {\bibfnamefont {I.}~\bibnamefont {Wygnanski}},\ }\bibfield
  {title} {\enquote {\bibinfo {title} {Large-scale structures in a forced
  turbulent mixing layer},}\ }\href@noop {} {\bibfield  {journal} {\bibinfo
  {journal} {Journal of Fluid Mechanics}\ }\textbf {\bibinfo {volume} {150}},\
  \bibinfo {pages} {23--39} (\bibinfo {year} {1985})}\BibitemShut {NoStop}%
\bibitem [{\citenamefont {Monkewitz}(1988)}]{monkewitz1988subharmonic}%
  \BibitemOpen
  \bibfield  {author} {\bibinfo {author} {\bibfnamefont {P.~A.}\ \bibnamefont
  {Monkewitz}},\ }\bibfield  {title} {\enquote {\bibinfo {title} {Subharmonic
  resonance, pairing and shredding in the mixing layer},}\ }\href@noop {}
  {\bibfield  {journal} {\bibinfo  {journal} {Journal of Fluid Mechanics}\
  }\textbf {\bibinfo {volume} {188}},\ \bibinfo {pages} {223--252} (\bibinfo
  {year} {1988})}\BibitemShut {NoStop}%
\bibitem [{\citenamefont {Ho}\ and\ \citenamefont
  {Huerre}(1984)}]{ho1984perturbed}%
  \BibitemOpen
  \bibfield  {author} {\bibinfo {author} {\bibfnamefont {C.-M.}\ \bibnamefont
  {Ho}}\ and\ \bibinfo {author} {\bibfnamefont {P.}~\bibnamefont {Huerre}},\
  }\bibfield  {title} {\enquote {\bibinfo {title} {Perturbed free shear
  layers},}\ }\href@noop {} {\bibfield  {journal} {\bibinfo  {journal} {Annual
  review of fluid mechanics}\ }\textbf {\bibinfo {volume} {16}},\ \bibinfo
  {pages} {365--422} (\bibinfo {year} {1984})}\BibitemShut {NoStop}%
\bibitem [{\citenamefont {Hussain}(1983)}]{hussain1983coherent}%
  \BibitemOpen
  \bibfield  {author} {\bibinfo {author} {\bibfnamefont {A.~F.}\ \bibnamefont
  {Hussain}},\ }\bibfield  {title} {\enquote {\bibinfo {title} {Coherent
  structures - reality and myth},}\ }\href@noop {} {\bibfield  {journal}
  {\bibinfo  {journal} {The Physics of fluids}\ }\textbf {\bibinfo {volume}
  {26}},\ \bibinfo {pages} {2816--2850} (\bibinfo {year} {1983})}\BibitemShut
  {NoStop}%
\bibitem [{\citenamefont {Husain}\ and\ \citenamefont
  {Hussain}(1995)}]{husain1995experiments}%
  \BibitemOpen
  \bibfield  {author} {\bibinfo {author} {\bibfnamefont {H.~S.}\ \bibnamefont
  {Husain}}\ and\ \bibinfo {author} {\bibfnamefont {F.}~\bibnamefont
  {Hussain}},\ }\bibfield  {title} {\enquote {\bibinfo {title} {Experiments on
  subharmonic resonance in a shear layer},}\ }\href@noop {} {\bibfield
  {journal} {\bibinfo  {journal} {Journal of Fluid Mechanics}\ }\textbf
  {\bibinfo {volume} {304}},\ \bibinfo {pages} {343--372} (\bibinfo {year}
  {1995})}\BibitemShut {NoStop}%
\bibitem [{\citenamefont {Morris}, \citenamefont {Giridharan},\ and\
  \citenamefont {Lilley}(1990)}]{morris1990turbulent}%
  \BibitemOpen
  \bibfield  {author} {\bibinfo {author} {\bibfnamefont {P.~J.}\ \bibnamefont
  {Morris}}, \bibinfo {author} {\bibfnamefont {M.~G.}\ \bibnamefont
  {Giridharan}}, \ and\ \bibinfo {author} {\bibfnamefont {G.~M.}\ \bibnamefont
  {Lilley}},\ }\bibfield  {title} {\enquote {\bibinfo {title} {On the turbulent
  mixing of compressible free shear layers},}\ }\href@noop {} {\bibfield
  {journal} {\bibinfo  {journal} {Proceedings of the Royal Society of London.
  Series A: Mathematical and Physical Sciences}\ }\textbf {\bibinfo {volume}
  {431}},\ \bibinfo {pages} {219--243} (\bibinfo {year} {1990})}\BibitemShut
  {NoStop}%
\bibitem [{\citenamefont {D'Ovidio}\ and\ \citenamefont
  {Coats}(2013)}]{d2013organized}%
  \BibitemOpen
  \bibfield  {author} {\bibinfo {author} {\bibfnamefont {A.}~\bibnamefont
  {D'Ovidio}}\ and\ \bibinfo {author} {\bibfnamefont {C.}~\bibnamefont
  {Coats}},\ }\bibfield  {title} {\enquote {\bibinfo {title} {Organized large
  structure in the post-transition mixing layer. {P}art 1. {E}xperimental
  evidence},}\ }\href@noop {} {\bibfield  {journal} {\bibinfo  {journal}
  {Journal of fluid mechanics}\ }\textbf {\bibinfo {volume} {737}},\ \bibinfo
  {pages} {466--498} (\bibinfo {year} {2013})}\BibitemShut {NoStop}%
\bibitem [{\citenamefont {Suryanarayanan}\ and\ \citenamefont
  {Narasimha}(2017)}]{suryanarayanan2017insights}%
  \BibitemOpen
  \bibfield  {author} {\bibinfo {author} {\bibfnamefont {S.}~\bibnamefont
  {Suryanarayanan}}\ and\ \bibinfo {author} {\bibfnamefont {R.}~\bibnamefont
  {Narasimha}},\ }\bibfield  {title} {\enquote {\bibinfo {title} {Insights into
  the growth rate of spatially evolving plane turbulent free-shear layers from
  2{D} vortex-gas simulations},}\ }\href@noop {} {\bibfield  {journal}
  {\bibinfo  {journal} {Physics of Fluids}\ }\textbf {\bibinfo {volume} {29}},\
  \bibinfo {pages} {020708} (\bibinfo {year} {2017})}\BibitemShut {NoStop}%
\bibitem [{\citenamefont {Brown}\ and\ \citenamefont
  {Roshko}(2012)}]{brown2012turbulent}%
  \BibitemOpen
  \bibfield  {author} {\bibinfo {author} {\bibfnamefont {G.~L.}\ \bibnamefont
  {Brown}}\ and\ \bibinfo {author} {\bibfnamefont {A.}~\bibnamefont {Roshko}},\
  }\bibfield  {title} {\enquote {\bibinfo {title} {Turbulent shear layers and
  wakes},}\ }\href@noop {} {\bibfield  {journal} {\bibinfo  {journal} {Journal
  of Turbulence}\ ,\ \bibinfo {pages} {N51}} (\bibinfo {year}
  {2012})}\BibitemShut {NoStop}%
\bibitem [{\citenamefont {Konrad}(1977)}]{konrad1977experimental}%
  \BibitemOpen
  \bibfield  {author} {\bibinfo {author} {\bibfnamefont {J.~H.}\ \bibnamefont
  {Konrad}},\ }\emph {\bibinfo {title} {An experimental investigation of mixing
  in two-dimensional turbulent shear flows with applications to
  diffusion-limited chemical reactions}},\ \href@noop {} {Ph.D. thesis},\
  \bibinfo  {school} {California Institute of Technology} (\bibinfo {year}
  {1977})\BibitemShut {NoStop}%
\bibitem [{\citenamefont {Dimotakis}(2000)}]{dimotakis2000mixing}%
  \BibitemOpen
  \bibfield  {author} {\bibinfo {author} {\bibfnamefont {P.~E.}\ \bibnamefont
  {Dimotakis}},\ }\bibfield  {title} {\enquote {\bibinfo {title} {The mixing
  transition in turbulent flows},}\ }\href@noop {} {\bibfield  {journal}
  {\bibinfo  {journal} {Journal of Fluid Mechanics}\ }\textbf {\bibinfo
  {volume} {409}},\ \bibinfo {pages} {69--98} (\bibinfo {year}
  {2000})}\BibitemShut {NoStop}%
\bibitem [{\citenamefont {Oster}\ and\ \citenamefont
  {Wygnanski}(1982)}]{oster1982forced}%
  \BibitemOpen
  \bibfield  {author} {\bibinfo {author} {\bibfnamefont {D.}~\bibnamefont
  {Oster}}\ and\ \bibinfo {author} {\bibfnamefont {I.}~\bibnamefont
  {Wygnanski}},\ }\bibfield  {title} {\enquote {\bibinfo {title} {The forced
  mixing layer between parallel streams},}\ }\href@noop {} {\bibfield
  {journal} {\bibinfo  {journal} {Journal of Fluid Mechanics}\ }\textbf
  {\bibinfo {volume} {123}},\ \bibinfo {pages} {91--130} (\bibinfo {year}
  {1982})}\BibitemShut {NoStop}%
\bibitem [{\citenamefont {Narasimha}(1990)}]{narasimha1990utility}%
  \BibitemOpen
  \bibfield  {author} {\bibinfo {author} {\bibfnamefont {R.}~\bibnamefont
  {Narasimha}},\ }\bibfield  {title} {\enquote {\bibinfo {title} {The utility
  and drawbacks of traditional approaches},}\ }in\ \href@noop {} {\emph
  {\bibinfo {booktitle} {Whither Turbulence? Turbulence at the Crossroads}}}\
  (\bibinfo  {publisher} {Springer},\ \bibinfo {year} {1990})\ pp.\ \bibinfo
  {pages} {13--48}\BibitemShut {NoStop}%
\bibitem [{\citenamefont {George}\ and\ \citenamefont
  {Davidson}(2004)}]{george2004role}%
  \BibitemOpen
  \bibfield  {author} {\bibinfo {author} {\bibfnamefont {W.~K.}\ \bibnamefont
  {George}}\ and\ \bibinfo {author} {\bibfnamefont {L.}~\bibnamefont
  {Davidson}},\ }\bibfield  {title} {\enquote {\bibinfo {title} {Role of
  initial conditions in establishing asymptotic flow behavior},}\ }\href@noop
  {} {\bibfield  {journal} {\bibinfo  {journal} {AIAA journal}\ }\textbf
  {\bibinfo {volume} {42}},\ \bibinfo {pages} {438--446} (\bibinfo {year}
  {2004})}\BibitemShut {NoStop}%
\bibitem [{\citenamefont {Suryanarayanan}, \citenamefont {Narasimha},\ and\
  \citenamefont {Dass}(2014)}]{snh}%
  \BibitemOpen
  \bibfield  {author} {\bibinfo {author} {\bibfnamefont {S.}~\bibnamefont
  {Suryanarayanan}}, \bibinfo {author} {\bibfnamefont {R.}~\bibnamefont
  {Narasimha}}, \ and\ \bibinfo {author} {\bibfnamefont {N.~H.}\ \bibnamefont
  {Dass}},\ }\bibfield  {title} {\enquote {\bibinfo {title} {Free turbulent
  shear layer in a point vortex gas as a problem in nonequilibrium statistical
  mechanics},}\ }\href@noop {} {\bibfield  {journal} {\bibinfo  {journal}
  {Physical Review E}\ }\textbf {\bibinfo {volume} {89}},\ \bibinfo {pages}
  {013009} (\bibinfo {year} {2014})}\BibitemShut {NoStop}%
\bibitem [{\citenamefont {Joyce}\ and\ \citenamefont
  {Montgomery}(1973)}]{joyce1973negative}%
  \BibitemOpen
  \bibfield  {author} {\bibinfo {author} {\bibfnamefont {G.}~\bibnamefont
  {Joyce}}\ and\ \bibinfo {author} {\bibfnamefont {D.}~\bibnamefont
  {Montgomery}},\ }\bibfield  {title} {\enquote {\bibinfo {title} {Negative
  temperature states for the two-dimensional guiding-centre plasma},}\
  }\href@noop {} {\bibfield  {journal} {\bibinfo  {journal} {Journal of Plasma
  Physics}\ }\textbf {\bibinfo {volume} {10}},\ \bibinfo {pages} {107--121}
  (\bibinfo {year} {1973})}\BibitemShut {NoStop}%
\bibitem [{\citenamefont {Spencer}\ and\ \citenamefont
  {Jones}(1971)}]{spencer1971statistical}%
  \BibitemOpen
  \bibfield  {author} {\bibinfo {author} {\bibfnamefont {B.}~\bibnamefont
  {Spencer}}\ and\ \bibinfo {author} {\bibfnamefont {B.}~\bibnamefont
  {Jones}},\ }\href@noop {} {\emph {\bibinfo {title} {Statistical investigation
  of pressure and velocity fields in the turbulent two-stream mixing layer}}}\
  (\bibinfo  {publisher} {American Institute of Aeronautics and Astronautics},\
  \bibinfo {year} {1971})\BibitemShut {NoStop}%
\bibitem [{\citenamefont
  {Krasny}(1986{\natexlab{a}})}]{krasny1986desingularization}%
  \BibitemOpen
  \bibfield  {author} {\bibinfo {author} {\bibfnamefont {R.}~\bibnamefont
  {Krasny}},\ }\bibfield  {title} {\enquote {\bibinfo {title}
  {Desingularization of periodic vortex sheet roll-up},}\ }\href@noop {}
  {\bibfield  {journal} {\bibinfo  {journal} {Journal of Computational
  Physics}\ }\textbf {\bibinfo {volume} {65}},\ \bibinfo {pages} {292--313}
  (\bibinfo {year} {1986}{\natexlab{a}})}\BibitemShut {NoStop}%
\bibitem [{\citenamefont {Cottet}, \citenamefont {Koumoutsakos}\ \emph
  {et~al.}(2000)\citenamefont {Cottet}, \citenamefont {Koumoutsakos} \emph
  {et~al.}}]{cottet2000vortex}%
  \BibitemOpen
  \bibfield  {author} {\bibinfo {author} {\bibfnamefont {G.-H.}\ \bibnamefont
  {Cottet}}, \bibinfo {author} {\bibfnamefont {P.~D.}\ \bibnamefont
  {Koumoutsakos}},  \emph {et~al.},\ }\href@noop {} {\emph {\bibinfo {title}
  {Vortex methods: theory and practice}}},\ Vol.~\bibinfo {volume} {8}\
  (\bibinfo  {publisher} {Cambridge university press Cambridge},\ \bibinfo
  {year} {2000})\BibitemShut {NoStop}%
\bibitem [{\citenamefont {Beale}\ and\ \citenamefont
  {Majda}(1982)}]{beale1982vortex}%
  \BibitemOpen
  \bibfield  {author} {\bibinfo {author} {\bibfnamefont {J.~T.}\ \bibnamefont
  {Beale}}\ and\ \bibinfo {author} {\bibfnamefont {A.}~\bibnamefont {Majda}},\
  }\bibfield  {title} {\enquote {\bibinfo {title} {Vortex methods. ii. higher
  order accuracy in two and three dimensions},}\ }\href@noop {} {\bibfield
  {journal} {\bibinfo  {journal} {Mathematics of Computation}\ }\textbf
  {\bibinfo {volume} {39}},\ \bibinfo {pages} {29--52} (\bibinfo {year}
  {1982})}\BibitemShut {NoStop}%
\bibitem [{\citenamefont {Marchioro}\ and\ \citenamefont
  {Pulvirenti}(1993)}]{marchioro1993}%
  \BibitemOpen
  \bibfield  {author} {\bibinfo {author} {\bibfnamefont {C.}~\bibnamefont
  {Marchioro}}\ and\ \bibinfo {author} {\bibfnamefont {M.}~\bibnamefont
  {Pulvirenti}},\ }\href@noop {} {\emph {\bibinfo {title} {Mathematical theory
  of incompressible nonviscous fluids}}},\ Vol.~\bibinfo {volume} {96}\
  (\bibinfo  {publisher} {Springer},\ \bibinfo {year} {1993})\BibitemShut
  {NoStop}%
\bibitem [{\citenamefont {Rosenhead}(1931)}]{rosenhead}%
  \BibitemOpen
  \bibfield  {author} {\bibinfo {author} {\bibfnamefont {L.}~\bibnamefont
  {Rosenhead}},\ }\bibfield  {title} {\enquote {\bibinfo {title} {The formation
  of vortices from a surface of discontinuity},}\ }\href@noop {} {\bibfield
  {journal} {\bibinfo  {journal} {Proceedings of the Royal Society of London.
  Series A}\ }\textbf {\bibinfo {volume} {134}},\ \bibinfo {pages} {170--192}
  (\bibinfo {year} {1931})}\BibitemShut {NoStop}%
\bibitem [{\citenamefont {Hama}\ and\ \citenamefont
  {Burke}(1960)}]{hama1960rolling}%
  \BibitemOpen
  \bibfield  {author} {\bibinfo {author} {\bibfnamefont {F.~R.}\ \bibnamefont
  {Hama}}\ and\ \bibinfo {author} {\bibfnamefont {E.~R.}\ \bibnamefont
  {Burke}},\ }\href@noop {} {\emph {\bibinfo {title} {On the rolling-up of a
  vortex sheet}}}\ (\bibinfo  {publisher} {University of Maryland, Institute
  for Fluid Dynamics and Applied Mathematics},\ \bibinfo {year}
  {1960})\BibitemShut {NoStop}%
\bibitem [{\citenamefont {Acton}(1976)}]{acton}%
  \BibitemOpen
  \bibfield  {author} {\bibinfo {author} {\bibfnamefont {E.}~\bibnamefont
  {Acton}},\ }\bibfield  {title} {\enquote {\bibinfo {title} {The modelling of
  large eddies in a two-dimensional shear layer},}\ }\href@noop {} {\bibfield
  {journal} {\bibinfo  {journal} {Journal of Fluid Mechanics}\ }\textbf
  {\bibinfo {volume} {76}},\ \bibinfo {pages} {561--592} (\bibinfo {year}
  {1976})}\BibitemShut {NoStop}%
\bibitem [{\citenamefont {Delcourt}\ and\ \citenamefont
  {Brown}(1979)}]{delcourtbrown}%
  \BibitemOpen
  \bibfield  {author} {\bibinfo {author} {\bibfnamefont {B.}~\bibnamefont
  {Delcourt}}\ and\ \bibinfo {author} {\bibfnamefont {G.}~\bibnamefont
  {Brown}},\ }\bibfield  {title} {\enquote {\bibinfo {title} {The evolution and
  emerging structure of a vortex sheet in an inviscid and viscous fluid
  modelled by a point vortex method},}\ }in\ \href@noop {} {\emph {\bibinfo
  {booktitle} {2nd Symposium on Turbulent Shear Flows}}},\ Vol.~\bibinfo
  {volume} {1}\ (\bibinfo {year} {1979})\ p.~\bibinfo {pages} {14}\BibitemShut
  {NoStop}%
\bibitem [{\citenamefont {Aref}\ and\ \citenamefont
  {Siggia}(1980)}]{arefsiggia}%
  \BibitemOpen
  \bibfield  {author} {\bibinfo {author} {\bibfnamefont {H.}~\bibnamefont
  {Aref}}\ and\ \bibinfo {author} {\bibfnamefont {E.}~\bibnamefont {Siggia}},\
  }\bibfield  {title} {\enquote {\bibinfo {title} {Vortex dynamics of the
  two-dimensional turbulent shear layer},}\ }\href@noop {} {\bibfield
  {journal} {\bibinfo  {journal} {J. Fluid Mech}\ }\textbf {\bibinfo {volume}
  {100}},\ \bibinfo {pages} {705--737} (\bibinfo {year} {1980})}\BibitemShut
  {NoStop}%
\bibitem [{\citenamefont {Lamb}(1932)}]{lamb1932hydrodynamics}%
  \BibitemOpen
  \bibfield  {author} {\bibinfo {author} {\bibfnamefont {H.}~\bibnamefont
  {Lamb}},\ }\href@noop {} {\enquote {\bibinfo {title} {Hydrodynamics cambridge
  university press},}\ } (\bibinfo {year} {1932})\BibitemShut {NoStop}%
\bibitem [{\citenamefont {v.~K{\'a}rm{\'a}n}\ and\ \citenamefont
  {Rubach}(1912)}]{v1912mechanismus}%
  \BibitemOpen
  \bibfield  {author} {\bibinfo {author} {\bibfnamefont {T.}~\bibnamefont
  {v.~K{\'a}rm{\'a}n}}\ and\ \bibinfo {author} {\bibfnamefont {H.}~\bibnamefont
  {Rubach}},\ }\bibfield  {title} {\enquote {\bibinfo {title} {{\"U}ber den
  mechanismus des fl{\"u}ssigkeits-und luftwiderstandes},}\ }\href@noop {}
  {\bibfield  {journal} {\bibinfo  {journal} {Phys. Zeitschr.}\ }\textbf
  {\bibinfo {volume} {13}},\ \bibinfo {pages} {49} (\bibinfo {year}
  {1912})}\BibitemShut {NoStop}%
\bibitem [{\citenamefont {Winant}\ and\ \citenamefont
  {Browand}(1974)}]{winant1974vortex}%
  \BibitemOpen
  \bibfield  {author} {\bibinfo {author} {\bibfnamefont {C.~D.}\ \bibnamefont
  {Winant}}\ and\ \bibinfo {author} {\bibfnamefont {F.~K.}\ \bibnamefont
  {Browand}},\ }\bibfield  {title} {\enquote {\bibinfo {title} {Vortex pairing:
  the mechanism of turbulent mixing-layer growth at moderate reynolds
  number},}\ }\href@noop {} {\bibfield  {journal} {\bibinfo  {journal} {Journal
  of Fluid Mechanics}\ }\textbf {\bibinfo {volume} {63}},\ \bibinfo {pages}
  {237--255} (\bibinfo {year} {1974})}\BibitemShut {NoStop}%
\bibitem [{\citenamefont {Chorin}(1973)}]{chorin1973numerical}%
  \BibitemOpen
  \bibfield  {author} {\bibinfo {author} {\bibfnamefont {A.~J.}\ \bibnamefont
  {Chorin}},\ }\bibfield  {title} {\enquote {\bibinfo {title} {Numerical study
  of slightly viscous flow},}\ }\href@noop {} {\bibfield  {journal} {\bibinfo
  {journal} {Journal of fluid mechanics}\ }\textbf {\bibinfo {volume} {57}},\
  \bibinfo {pages} {785--796} (\bibinfo {year} {1973})}\BibitemShut {NoStop}%
\bibitem [{\citenamefont {Dutta}(1988)}]{dutta1988discrete}%
  \BibitemOpen
  \bibfield  {author} {\bibinfo {author} {\bibfnamefont {P.}~\bibnamefont
  {Dutta}},\ }\emph {\bibinfo {title} {Discrete vortex method for separated and
  free shear flows}},\ \href@noop {} {Ph.D. thesis},\ \bibinfo  {school} {Ph.
  D. Thesis, Indian Institute of Science, Bangalore} (\bibinfo {year}
  {1988})\BibitemShut {NoStop}%
\bibitem [{\citenamefont {Suryanarayanan}\ and\ \citenamefont
  {Brown}(2011)}]{SuryanarayananBrown2011}%
  \BibitemOpen
  \bibfield  {author} {\bibinfo {author} {\bibfnamefont {S.}~\bibnamefont
  {Suryanarayanan}}\ and\ \bibinfo {author} {\bibfnamefont {G.~L.}\
  \bibnamefont {Brown}},\ }\bibfield  {title} {\enquote {\bibinfo {title}
  {Linear and non-linear stability via {B}iot-{S}avart computations},}\ }in\
  \href@noop {} {\emph {\bibinfo {booktitle} {IUTAM Symposium on Bluff Body
  Flows, IIT-Kanpur, India}}}\ (\bibinfo {year} {2011})\ pp.\ \bibinfo {pages}
  {63--66}\BibitemShut {NoStop}%
\bibitem [{\citenamefont {Aref}(1983)}]{aref1983integrable}%
  \BibitemOpen
  \bibfield  {author} {\bibinfo {author} {\bibfnamefont {H.}~\bibnamefont
  {Aref}},\ }\bibfield  {title} {\enquote {\bibinfo {title} {Integrable,
  chaotic, and turbulent vortex motion in two-dimensional flows},}\ }\href@noop
  {} {\bibfield  {journal} {\bibinfo  {journal} {Annual Review of Fluid
  Mechanics}\ }\textbf {\bibinfo {volume} {15}},\ \bibinfo {pages} {345--389}
  (\bibinfo {year} {1983})}\BibitemShut {NoStop}%
\bibitem [{\citenamefont {Novikov}\ and\ \citenamefont
  {Sedov}(1978)}]{novikov1978stochastic}%
  \BibitemOpen
  \bibfield  {author} {\bibinfo {author} {\bibfnamefont {E.}~\bibnamefont
  {Novikov}}\ and\ \bibinfo {author} {\bibfnamefont {Y.~B.}\ \bibnamefont
  {Sedov}},\ }\bibfield  {title} {\enquote {\bibinfo {title} {Stochastic
  properties of a four-vortex system},}\ }\href@noop {} {\bibfield  {journal}
  {\bibinfo  {journal} {Z. Eksp. Teor. Fiz}\ }\textbf {\bibinfo {volume}
  {75}},\ \bibinfo {pages} {868--876} (\bibinfo {year} {1978})}\BibitemShut
  {NoStop}%
\bibitem [{\citenamefont {Birkhoff}(1962)}]{birkhoff1962helmholtz}%
  \BibitemOpen
  \bibfield  {author} {\bibinfo {author} {\bibfnamefont {G.}~\bibnamefont
  {Birkhoff}},\ }\bibfield  {title} {\enquote {\bibinfo {title} {Helmholtz and
  {T}aylor instability},}\ }in\ \href@noop {} {\emph {\bibinfo {booktitle}
  {Proc. Symp. Appl. Math}}},\ Vol.~\bibinfo {volume} {13}\ (\bibinfo {year}
  {1962})\ pp.\ \bibinfo {pages} {55--76}\BibitemShut {NoStop}%
\bibitem [{\citenamefont {Moore}(1979)}]{moore1979spontaneous}%
  \BibitemOpen
  \bibfield  {author} {\bibinfo {author} {\bibfnamefont {D.~W.}\ \bibnamefont
  {Moore}},\ }\bibfield  {title} {\enquote {\bibinfo {title} {The spontaneous
  appearance of a singularity in the shape of an evolving vortex sheet},}\
  }\href@noop {} {\bibfield  {journal} {\bibinfo  {journal} {Proceedings of the
  Royal Society of London. A. Mathematical and Physical Sciences}\ }\textbf
  {\bibinfo {volume} {365}},\ \bibinfo {pages} {105--119} (\bibinfo {year}
  {1979})}\BibitemShut {NoStop}%
\bibitem [{\citenamefont {Krasny}(1986{\natexlab{b}})}]{krasny1986study}%
  \BibitemOpen
  \bibfield  {author} {\bibinfo {author} {\bibfnamefont {R.}~\bibnamefont
  {Krasny}},\ }\bibfield  {title} {\enquote {\bibinfo {title} {A study of
  singularity formation in a vortex sheet by the point-vortex approximation},}\
  }\href@noop {} {\bibfield  {journal} {\bibinfo  {journal} {Journal of Fluid
  Mechanics}\ }\textbf {\bibinfo {volume} {167}},\ \bibinfo {pages} {65--93}
  (\bibinfo {year} {1986}{\natexlab{b}})}\BibitemShut {NoStop}%
\bibitem [{\citenamefont {Hald}(1979)}]{hald1979convergence}%
  \BibitemOpen
  \bibfield  {author} {\bibinfo {author} {\bibfnamefont {O.~H.}\ \bibnamefont
  {Hald}},\ }\bibfield  {title} {\enquote {\bibinfo {title} {Convergence of
  vortex methods for {E}uler's equations. {II}},}\ }\href@noop {} {\bibfield
  {journal} {\bibinfo  {journal} {SIAM Journal on Numerical Analysis}\ }\textbf
  {\bibinfo {volume} {16}},\ \bibinfo {pages} {726--755} (\bibinfo {year}
  {1979})}\BibitemShut {NoStop}%
\bibitem [{\citenamefont {Goodman}, \citenamefont {Hou},\ and\ \citenamefont
  {Lowengrub}(1990)}]{goodman1990convergence}%
  \BibitemOpen
  \bibfield  {author} {\bibinfo {author} {\bibfnamefont {J.}~\bibnamefont
  {Goodman}}, \bibinfo {author} {\bibfnamefont {T.~Y.}\ \bibnamefont {Hou}}, \
  and\ \bibinfo {author} {\bibfnamefont {J.}~\bibnamefont {Lowengrub}},\
  }\bibfield  {title} {\enquote {\bibinfo {title} {Convergence of the point
  vortex method for the 2-{D} euler equations},}\ }\href@noop {} {\bibfield
  {journal} {\bibinfo  {journal} {Communications on Pure and Applied
  Mathematics}\ }\textbf {\bibinfo {volume} {43}},\ \bibinfo {pages} {415--430}
  (\bibinfo {year} {1990})}\BibitemShut {NoStop}%
\bibitem [{\citenamefont {Suryanarayanan}(2015)}]{suryanarayanan2014insights}%
  \BibitemOpen
  \bibfield  {author} {\bibinfo {author} {\bibfnamefont {S.}~\bibnamefont
  {Suryanarayanan}},\ }\emph {\bibinfo {title} {Insights into turbulent
  free-shear-layer dynamics from vortex-gas computations and statistical
  mechanics}},\ \href@noop {} {Ph.D. thesis},\ \bibinfo  {school} {Jawaharlal
  Nehru Centre for Advanced Scientific Research} (\bibinfo {year}
  {2015})\BibitemShut {NoStop}%
\end{thebibliography}%

\end{document}